\documentclass[useAMS,usenatbib]{mn2e}
\usepackage[fleqn]{amsmath}
\usepackage{graphicx}
\usepackage{color}

\title[Evolution of galaxies with and without bulges]{The evolution of disc galaxies with and without classical bulges since $z\sim1$}
\author[Sonali Sachdeva, Dimitri A. Gadotti, Kanak Saha, Harinder P. Singh]{Sonali Sachdeva$^{1}$, Dimitri A. Gadotti$^{2}$, Kanak Saha$^{3}$, Harinder P. Singh$^{1}$\\
$^{1}$Department of Physics and Astrophysics, University of Delhi, Delhi 110007, India\\
$^{2}$European Southern Observatory, Alonso de Cordova 3107, Vitacura, Casilla 19001, Santiago, Chile\\
$^{3}$Inter-University Centre for Astronomy and Astrophysics, Pune 411007, India}
\begin{document}

\date{in original form 2014 December 30}

\pagerange{\pageref{firstpage}--\pageref{lastpage}} \pubyear{2014}

\maketitle

\label{firstpage}

\begin{abstract}
Establishing the relative role of internally and externally driven mechanisms responsible for disc and bulge growth is essential to understand the evolution of disc galaxies. In this context, we have studied the physical properties of disc galaxies without classical bulges in comparison to those with classical bulges since $z\sim0.9$. Using images from the {\it Hubble Space Telescope} and {\it Sloan Digital Sky Survey}, we have computed both parametric and non-parametric measures, and examined the evolution in size, concentration, stellar mass, effective stellar mass density and asymmetry. We find that both disc galaxies with and without classical bulges have gained more than 50\% of their present stellar mass over the last $\sim$8 Gyrs. Also, the increase in disc size is found to be peripheral. While the average total (Petrosian) radius almost doubles from $z\sim0.9$ to $z\sim0$, the average effective radius undergoes a marginal increase in comparison. Additionally, increase in the density of the inner region is evident through the evolution of both concentration and effective stellar mass density. We find that the asymmetry index falls from higher to lower redshifts, but this is more pronounced for the bulgeless disc sample. Also, asymmetry correlates with the global effective radius, and concentration correlates with the global S\'ersic index, but better so for higher redshifts only. The substantial increase in mass and size indicates that accretion of external material has been a dominant mode of galaxy growth, where the circumgalactic environment plays a significant role.
\end{abstract}

\begin{keywords}
galaxies: bulges -- galaxies: evolution -- galaxies: high-redshift -- galaxies: structure.
\end{keywords}


\section{Introduction}
One of the major challenges linked to the morphological evolution of disc galaxies is to understand the formation and evolution of their bulge component. The bulge of a disc galaxy, observationally, is the central component which contains all the light that is in excess of an inward extrapolation of a constant scale-length exponential disk \citep{Wyseetal1997,Buta2013}. Internal and external mechanisms can explain the growth of bulges in disc galaxies.

Externally driven bulge growth in discs can occur through major mergers, multiple minor mergers, as well as accretion of small components or satellites \citep{Kauffmannetal1993,Baughetal1996,Aguerrietal2001,Bournaudetal2007,Hopkinsetal2010}. The fraction of baryons which lose angular-momentum due to the dynamical friction produced in these mergers and accretion processes fall to the centre of the galaxy to form the bulge component \citep[][see also recent review by \citealt{BrooksandChristensen2015}]{Parryetal2009,Governatoetal2010,Brooketal2011}.

In terms of internally driven, there can be two bulge-building mechanisms. The first one is known as secular evolution. In this mechanism, slow, internally created processes contribute to the rearrangement of angular momentum and mass inside disc galaxies, leading to the growth of pseudo-bulges, which have properties similar to discs \citep{BinneyandTremaine1987,Combes2001,KormendyandKennicutt2004,Kormendyetal2010,SahaandGerhard2013}. In the second mechanism, the coalescence of giant star forming clumps, due to internal gravitational instabilities, leads to bulge formation in discs \citep[][see recent review by \citealt{Bournaud2015}]{Elmegreenetal2008}. The relative importance of these mechanisms in forming present day discs with varied bulge properties is as yet not quantitatively known \citep[see discussion in][]{Kormendy2015}. 

One of the most prominent unresolved issues associated with bulge evolution is the conspicuous presence of a large number of bulgeless galaxies (disc galaxies without classical bulges) in the local universe. This is because, naively, a relatively low number of disc galaxies, formed as per the $\Lambda$CDM structure formation scenario, are expected to survive without forming a classical bulge in the centre. The intense amount of merger violence associated with hierarchical clustering is expected to destroy the fragile thin disc of stars. Thus, their presence is a huge challenge as emphasized through observations \citep{Kautschetal2006,Kautsch2009,Kormendyetal2010}, models \citep{Hopkinsetal2009,Governatoetal2010}, simulations \citep{Scannapiecoetal2009,Zavalaetal2012} and detailed in reviews \citep{Baugh2006,Benson2010,PeeblesandNusser2010,Kormendy2013}.

Observational evidence is required to pin down the relative role of various mechanisms in bulge growth and understand the formation of discs without bulges. The constraint involved is that for distant galaxies only the very basics of structure and morphology can be investigated. However, recent developments in observational facilities and in parametric and non-parametric measurements of galaxy structure \citep[][and references therein]{Conselice2014} have enabled us to use galaxy structure to measure fundamental properties of distant galaxies. These properties can then be compared with those of nearby galaxies to determine bulge-disc evolution.

There has been some progress in this regard. \citet{Gadotti2009} found an overlap in the structural properties of pseudo- and classical bulges indicating that the different processes for bulge growth might have happened concomitantly. \citet{Parryetal2009} studied the models based on $\Lambda$CDM cosmology and found that most spiral bulges acquire their stellar mass through minor mergers and disc instabilities. \citet{Moetal2010} explained the formation of bulges through mergers, secular internal processes as well as misaligned/perturbed infalling gas. \citet{Watsonetal2011} studied the neutral hydrogen properties of 20 bulgeless galaxies to compare the role of mergers vs internal processes in bulge formation. They report that even though some of the discs have distinct outer components indicating recent interaction, the discs remain bulgeless. \citet{Zavalaetal2012} examined the impact of mergers on the growth of bulges using simulated data. They found that the main channels of bulge mass assembly are stars from infalling satellites, and stars transferred from primary discs due to merger induced perturbations. \citet{Perezetal2013} showed through simulations that strong disc instabilities at high redshifts lead to classical bulge formation, which cannot be prevented by even the most energetic supernova feedback. \citet{Bruceetal2014} found that from redshift 3 to 1, galaxies move from disc dominated to increasingly bulge dominated morphology. 

Although these works have given us considerable insight into disc structural evolution, the relative role of internal and external mechanisms in the formation and evolution of discs of varied bulge types is as yet not established. To achieve this, it will be insightful to examine the evolution in the inner region properties of disc galaxies of different bulge types in a relative manner. The crucial aspect in that direction is that the separation of disc galaxies according to their bulge type has to be achieved in a quantitative and robust manner.

Thus, in this paper, we undertake a comparative study of the evolution of disc galaxies with and without classical bulges. Disc galaxies without classical bulges are, by definition, labelled as ``bulgeless'' and those with classical bulges are labelled as ``normal''. The two morphological types are separated using both S\'ersic index and Kormendy relation criteria \citep{Gadotti2009}. This work is a follow-up to \citet{Sachdeva2013}, such that the galaxy sample is the same and we also utilize the parametric measures (through S\'ersic function fitting) derived in that paper. Here we derive the non-parametric measures (Petrosian radius, concentration, asymmetry) along with rest-frame colours, total stellar mass and effective stellar mass density. The evolution in the inner region properties is thus examined through both parametric as well as non-parametric measures. 

The study is done since $z\sim1$, when the galaxies have just formed a familiar Hubble sequence structure \citep{Conseliceetal2011,Mortlocketal2013}, to the present epoch, where they have developed and settled into few distinctly identifiable categories \citep{Gadotti2009,Buta2013,Graham2013}. This time interval has the potential to reveal the major processes involved in bulge and disc evolution. 

For distant galaxies we make use of deep imaging from the Great Observatories Origins Deep Survey (GOODS) obtained using {\it Hubble Space Telescope (HST)}\footnote{Based on observations obtained with the NASA/ESA {\it Hubble Space Telescope}, which is operated by the Association of Universities for Research in Astronomy, Inc.(AURA) under NASA contract NAS 5-26555.}- Advanced Camera for Surveys in the Chandra Deep Field South (CDF-S) \citep{Giavaliscoetal2004}. For local galaxies, images are from the NASA Sloan Atlas\footnote{http://www.nsatlas.org. Funding for the NASA-Sloan Atlas has been provided by the NASA Astrophysics Data Analysis Program (08-ADP08-0072) and the NSF (AST-1211644).}, based on the Sloan Digital Sky Survey \citep{Blantonetal2005a}. We obtain the parametric and non-parametric measures for bright (M$_B$$\leq$-20) disc dominated (S\'ersic index $n<2.5$) galaxies in three redshift ranges (0.77$\leq z<$1.0, 0.4$\leq z<$0.77 and 0.02$\leq z<$0.05) for rest-frame {\it B}-band. We then examine the evolution of the inner-region properties of the bulgeless disc galaxies in comparison to the normal disc galaxies over the three redshift ranges. 

We consider a flat $\Lambda$-dominated universe with $\Omega_{\Lambda}=0.73$, $\Omega_m=0.27$, $H_o=71$ km sec$^{-1}$ Mpc$^{-1}$. In Section 2, we describe our data in terms of sample selection, preparation of images, defining and computation of various parameters. In Section 3, we present the results obtained by examining the evolution of size, concentration, stellar density and asymmetry. We also report the correlation seen between various parameters. In Section 4, we list the primary results of this work and discuss their implications in the light of previous studies.


\section{Data}

\subsection{S\'ersic parameters}
In a previous work \citep{Sachdeva2013}, images taken from HST - ACS in {\it V} (F606W), {\it i} (F775W), and {\it z} (F850LP) filters were used to obtain the rest-frame {\it B}-band properties of the galaxies lying in the CDF-S with redshift ranging from 0.4 to 1.0. First, SExtractor \citep{BertinandArnouts1996} was used to identify the sources in the {\it z}-band image. Then, single S\'ersic \citep{Sersic1968} components were fit \citep[using GALFIT,][]{Pengetal2002} on all the three filter images. The S\'ersic profile for the variation of a galaxy's surface brightness from its centre is given as:
\begin{equation} 
I(r)=I_e\exp[-b_n((\frac{r}{r_e})^{1/n}-1)],
\end{equation}
where $n$ (the S\'ersic index) controls the degree of curvature of the profile, $I_e$ is the surface brightness at $r_e$, and $b_n$ is a constant such that $r_e$ is the half-light radius for a given value of n. We thus obtained parameters such as apparent total magnitude, half-light radius and S\'ersic index for all the galaxies in the three filters.

Redshifts were obtained from the COMBO-17 survey \citep{Wolfetal2004}. The {\it V} filter provides rest-frame {\it B}-band properties for galaxies lying in the redshift range of 0.4 to 0.6. The {\it z} filter provides the same for the redshift range of 0.8 to 1.0. For the redshift range of 0.6 to 0.8, both {\it i} and {\it z} filters can be used and thus an average of the properties obtained using both the filters was employed. The parameters were thus obtained for 4124 sources in the rest-frame {\it B}-band for the full redshift range (0.4 to 1.0).

Using the redshifts and accepted cosmological parameters, we then computed absolute magnitude, half-light radius in kpc and surface brightness in mag/arcsec$^2$. The absolute-magnitude, M, for the galaxies is calculated using the relation:
\begin{equation}
M=m-5*log_{10}(D_L*10^5)-K,
\end{equation}
where $D_L$ is the luminosity-distance in Mpc and K is the K-correction term that accounts for the difference between the observed band and rest-frame band. It depends on the object's spectral energy distribution \citep{OkeandSandage1968,Hoggetal2002} and, for a power-law continuum, it is given by the relation:
\begin{equation}
K_{cont}=-2.5*(1+\alpha_{\nu})*\log(1+z),
\end{equation}
where $\alpha_{\nu}$ is the slope of the continuum and has a canonical value of -0.5 \citep{Richardsetal2006}. Therefore:
\begin{equation}
M_B=m-5*\log(D_L*10^5)+2.5*\log(\sqrt(1+z)).
\end{equation}

The half-light radius of the galaxies that we obtain from S\'ersic component fitting is in pixels. They were converted into arc-seconds according to the plate-scale of the telescope and then into radians. The intrinsic half-light radius in kpc was then calculated using the relation:
\begin{equation}
R_e=D_A*1000*\Delta\Theta,
\end{equation}
where $D_A$ is the angular-diameter-distance in Mpc and $\Delta\Theta$ is the radians covered on the detector by the half-light radius.

The redshift-magnitude distribution of the galaxies was examined in the 0.4-1.0 redshift range for galaxies with $M_B$$>$-20 and $M_B$$\leq$-20 \citep[shown in Fig.~1 of][]{Sachdeva2013}. For $M_B$$>$-20, galaxies with lower luminosities were not seen at high redshift ranges at all. However, for $M_B$$\leq$-20, the number of galaxies was seen to be evenly distributed. Also, the number of galaxies in two equal comoving volume redshift bins (0.4-0.77 and 0.77-1.0) was found to be almost the same for $M_B$$\leq$-20. Additionally, the magnitude limit for a reliable redshift estimate from COMBO-17 is $m_Z$$\sim$23.5, which for our upper redshift limit of $z=1.0$ corresponds to $M_B$$\sim$-20. Based on the depth of {\it HST} imaging, and the redshift accuracy limit of COMBO-17, a magnitude cut of -20 was applied on the sample. We, thus, obtained 727 sources in the rest-frame {\it B}-band (0.4$\leq z<$1.0) with $M_B$$\leq$-20 \citep[elaborated in][]{Sachdeva2013}.

\subsubsection{Separating bulgeless disc and normal disc galaxies} 
The S\'ersic index value of 2.5 is employed in numerous studies to separate early-type (n$>$2.5) and late-type (n$<$2.5) galaxies \citep[e.g.][]{Ravindranathetal2004,Bardenetal2005,vanderWel2008}. We used these criteria to obtain 496 late-type (or disc dominated) (n$<$2.5), bright ($M_B$$\leq$-20) galaxies in the rest-frame {\it B}-band.

The bulgeless disc galaxies include disc galaxies without bulges and those with pseudo-bulges. Pseudo-bulges are the bulges which have a higher ratio of ordered-motion to random-motion. Since they exhibit nearly exponential brightness profiles, disc galaxies with pseudo-bulges are, therefore, considered as bulgeless \citep{KormendyandKennicutt2004,Kormendyetal2010}. Thus, to separate bulgeless disc and normal disc galaxies, we separated discs with no-bulge or pseudo-bulge from discs with classical bulge in our disc dominated sample.

The S\'ersic index values ranging from 1.7-2.0 have been suggested by many studies for the separation of classical bulges from pseudo bulges \citep[e.g.][]{Shenetal2003,Laurikainenetal2007,FisherandDrory2008}. To obtain a S\'ersic index limit for our sample, we divided the entire sample (i.e. without the magnitude cut, 4124 sources) into three ranges (0.8$>n$, 0.8$\leq n<$1.7, 1.7$\leq n$) with each range getting almost equal number of sources. We then examined the distribution of the mean half-light radius against the absolute magnitude bins for different S\'ersic index ranges \citep[see Fig.~2 of][]{Sachdeva2013}.

We found that a value of n$\sim$1.7 divides galaxies into two groups where each group follows a particular half-light radius-magnitude ($\overline{R}$-M) relation, independent of n \citep[shown in Fig.~2 of][]{Sachdeva2013}. This is in striking agreement with \citet{Shenetal2003}, who found the same value based on SDSS data, and claimed that the cut separates galaxies with exponential surface brightness profile (Sb/Sc) from the galaxies which do not have such profiles.

In addition to applying the S\'ersic index limit of n$\sim$1.7, we have applied the Kormendy relation \citep{Kormendy1977} to ascertain the separation of bulgeless disc and normal disc galaxies. This is based on the fact that, since the Kormendy relation is followed by elliptical galaxies, galaxies which are bulgeless should show themselves as outliers to the relation \citep[see][for details]{Gadotti2009}.

To achieve this, we fitted a linear relation to the surface brightness vs log-size data of elliptical galaxies in our sample, i.e. those with $2.8\leq n<4.5$ from the 727 sources (0.4$\leq z<$1.0, $M_B$$\leq$-20). Those disc galaxies ($n<2.5$) which were lying below the $\pm$3$\sigma$ value of the zero-point (with fixed slope) were taken as outliers. The outliers obey the following relations: 
\begin{equation}
\mu_{e,B}>19.36+2.92*\log(R_{e,B}),0.4\leq z<0.77
\end{equation}
and
\begin{equation}
\mu_{e,B}>19.32+2.92*\log(R_{e,B}),0.77\leq z<1.0.
\end{equation}
More than 80$\%$ of the disc galaxies which were found to be bulgeless according to these relations were also seen to have n$<$1.7. The two criteria are thus complementary to each other. Only those galaxies which satisfied both criteria, i.e. had S\'ersic index less than 1.7 and were outliers to the Kormendy relation, were chosen to be bulgeless \citep[Fig.~3 of][]{Sachdeva2013}. The bulge/total light ratio was found to be less than 0.2 (or 20\%) for our bulgeless sample. This process of morphological determination was found to separate the galaxies in a similar manner in the infrared as they do in the optical \citep{Sachdeva2013}.

\subsubsection{Overall sample obtained}
We obtained S\'ersic parameters in rest-frame {\it B}-band for 496 bright, disc dominated galaxies separated into 186 bulgeless disc galaxies and 310 normal disc galaxies, in two equal comoving volume redshift ranges (0.77$\leq z<$1.0 and 0.4$\leq z<$0.77).

In addition to this, a low redshift (0.02$\leq z<$0.05) catalog of disc dominated galaxies was taken from the NYU-VAGC \citep{Blantonetal2005a,Blantonetal2005b} to establish a local sample. We obtained rest-frame {\it B}-band parameters from the single S\'ersic fit parameters in {\it g} and {\it r} filters. Their AB magnitudes are galactic-extinction corrected \citep{Schlegeletal1998} and also K-corrected \citep{Blantonetal2003} to the rest-frame bandpasses. We used relations from \citet{Fukugitaetal1996} and \citet{Jesteretal2005} to obtain the absolute magnitude in rest-frame {\it B}-band:
\begin{equation}
g^{\prime}=V+0.56*(B-V)-0.12,
\end{equation}
\begin{equation}
r^{\prime}=V-0.49*(B-V)+0.11.
\end{equation}
Using those two equations we get:
\begin{equation}
B=1.419*g^{\prime}-0.419*r^{\prime}+0.216.
\end{equation}
The S\'ersic half-light radii of the {\it g} and {\it r}-bands were converted from arc-seconds to kpc according to their redshift. The relations given by \citet{Bardenetal2005}, that were found using the \citet{deJong1996} data, were used to obtain half-light-radii in the rest-frame {\it B}-band. Using the NYU-VAGC catalog, they studied the ratio of half-light sizes in the five SDSS bands to the size measured in one band as a function of wavelength. The results were accurately described by a linear fit (only $\pm$3\% correction factor) and were in striking agreement with the fit found for the \citet{deJong1996} data. The slope gives average corrections to obtain the rest-frame sizes: 
\begin{equation}
R_e(V)=1.011*R_e(r),
\end{equation}
\begin{equation}
R_e(B)=1.017*R_e(V).
\end{equation}
Using those two we get:
\begin{equation}
R_e(B)=1.017*1.011*R_e(r).
\end{equation}

We obtained the final local catalog of 764 galaxies in rest-frame {\it B}-band, in the redshift-range 0.02$\leq$z$<$0.05 with $M_B$$\leq$-20.
After that we applied the S\'ersic index criteria to separate disc galaxies. Then disc galaxies were further separated into a sample of bulgeless disc and normal disc galaxies using both the S\'ersic index and Kormendy relation criteria, as described for the main sample.

Overall, we obtained S\'ersic parameters in rest-frame {\it B}-band for 597 bright disc galaxies separated into 211 bulgeless disc and 386 normal disc galaxies over three redshift ranges (0.77$\leq z<$1.0, 0.4$\leq z<$0.77 and 0.02$\leq z<$0.05). For details and the catalog please consult \citet{Sachdeva2013}.

\subsection{Obtaining and cleaning the images}
To probe the formation and evolution of bulges in disc galaxies with time, it is required to do image analysis and compute parameters like concentration and asymmetry of stellar light in each individual galaxy of this sample.

For the main sample (0.4$\leq z<$1.0), a 10 arcsec cutout is downloaded from the HST-ACS data archive and for the local sample (0.02$\leq z<$0.05), a 3 arcmin cutout is downloaded from the NASA-SLOAN ATLAS data archive. Since the aim is to do the study in rest-frame {\it B}-band, the filter chosen for obtaining the galaxy image is according to the redshift of the galaxy. The images are thus taken in {\it V}, {\it i}, {\it z} and {\it g} filters for redshifts 0.4-0.6, 0.6-0.8, 0.8-1.0 and 0.02-0.05 respectively. The cutout is such that the centre of the galaxy (brightest pixel) is at the centre of the image and an average-size galaxy covers not more than 60\% of the total area.

Out of 597 downloaded images, 27 source images are not taken up for analysis. This is because 12 of these galaxies are only partially imaged and the rest of the 15 galaxies have a highly fragmented light distribution which in part appears to be due to multiple overlapping sources. For the latter galaxies, it is difficult to determine and study their isolated light distribution. Also, since there is a lack of proper structure and there is a degeneracy of bright patches, there is no clear area from where the initial central pixel value can be selected which is crucial for the computation of radius, concentration and asymmetry. Out of these 15 sources, 3 are in the high redshift range (0.77-1.0) of the bulgeless disc category constituting $\sim$1\% of this sample. Five are in the middle redshift range (0.4-0.77) of the normal disc category constituting less than $\sim$4\% of this sample. The rest of the 15 are in the high redshift range (0.77-1.0) of the normal disc category and they also constitute $\sim$4\% of this sample. All these galaxies cover a large range of luminosity, S\'ersic index and half-light radius values. Their small fractions and even distribution in terms of redshift and parameter values indicate that their removal from the total sample should not affect the statistical estimations. 

The major task involved now is to clean or decontaminate the 570 galaxy images, i.e., to remove the neighbouring sources. These images are final in the sense of flat-fielding, bias subtraction, cosmic ray removal etc. To clean the images, each galaxy image is taken up separately and the value of the pixels covered by a neighbouring source is replaced with the average value of the sky pixels surrounding that source. The neighbouring sources are recognized through the use of SExtractor's catalog and segmentation map. Along with all the individual object coordinates, the SExtractor also provides an estimation of the radius of the object that contains more than 90\% of its light. The replacing of pixels is done using IRAF (Image Reduction and Analysis Facility) {\it imedit} task, which creates a circular annulus of a chosen radius around the selected central coordinates of the source and replaces the pixels within the inner circle with the average value of the pixels inside the annulus. We have tried to ensure that the masking process in each of the galaxy images does not strongly affect the light distribution of the outer parts of the main source.

Next important thing is to remove the background flux from all the galaxy images. To estimate the background flux, a blank patch (i.e. which even without masking is devoid of any light sources) is selected near the galaxy. The patch is chosen in such a way that it is small and far enough from the source, so as not to have any diffuse light, and yet large and near enough to give a reasonable estimate. The HST object mosaics have an area of 333$\times$333 pixels (10 arcsec squared, 0.03 arcsec/pixel) and the selected background patch in each mosaic covers an area of 30$\times$30 pixels. The SDSS object mosaics have an area of 440$\times$440 pixels (3 arcmin squared, 0.396 arcsec/pixel) and the selected background patch in each mosaic covers an area of 40$\times$40 pixels. The mean flux per pixel is then estimated from this patch and subtracted out from the image. The stability of the process is ascertained by selecting a large number of blank patches for some of the galaxy images. It is seen that they provide consistent background flux values. After cleaning and background flux subtraction, the images are ready for the computation of various parameters.

\subsection{Computing Petrosian radii} 
The total radius of a disc galaxy is difficult to be determined in a reproducible manner due to difficulties regarding its extent in terms of the dark matter halo. Also, to measure the optical or stellar extent, use of a fixed isophotal value of brightness is not optimal, due to the vast range of surface brightnesses in which galaxies exist. However, it is important to be determined in such a way that sizes and the associated parameters can be compared for galaxies of different luminosities and distances.

In that regard, the Petrosian radius \citep{Petrosian1976} is considered an effective way of measuring galaxy sizes \citep{Bershadyetal2000,Grahametal2005}. This radius is determined by tracking the ratio of surface brightness at each successively increasing radius to that averaged inside the radius. When the ratio ($\eta$$=$I(r)/$<$I($<$r)$>$) falls to a chosen small fraction (we take 0.2), that radius is multiplied by a factor (we take 1.5) to obtain the Petrosian radius. The values 0.2 and 1.5 are determined to be most appropriate \citep{Conselice2003,Lotzetal2004}. For our sample, we first determine the centre accurately using the IRAF task {\it imcentroid} and then measure the surface brightness at and inside successively increasing radii. Then, using 
\begin{equation}
\eta(r_p)=\frac{I(r_p)}{<I(<r_p)>}=0.2,
\end{equation}
\begin{equation}
R_P=1.5*(r_p),
\end{equation}
we measure R$_P$ (the Petrosian radius) in rest-frame {\it B}-band for all the galaxies. This radius is in pixels and converted into kpc according to the redshift and chosen cosmology.

\subsection{Computing Concentration}
For the well resolved low redshift galaxies, the bulge prominence is quantified and studied through e.g. bulge/disc decompositions \citep[e.g.][]{Gadotti2008,Simardetal2011}. However, for high redshifts, single/global S\'ersic indices are commonly used to obtain a quantitative assessment of the bulge component \citep[e.g.][and references therein]{Buitragoetal2013,Moslehetal2013}. We have global S\'ersic index values for all the galaxies in our sample, obtained earlier from S\'ersic function fitting \citep{Sachdeva2013}. We now compute concentration, which is considered a more robust parameter in terms of surface brightness dimming \citep[detailed in][]{Grahametal2005} and a better estimator of the bulge-to-total ratio \citep{Gadotti2009}.

To compute concentration, we first measure the total flux (or total counts) inside the Petrosian radius, i.e. the full flux from the source. Then, the number of counts in successively increasing radii, from the centre, are computed for each galaxy. When the number of counts is 20\% of the total number of counts, that radius is taken as the `inner-radius'. When it's 80\%, that radius is taken as the `outer-radius'. Concentration of the source is defined as \citep{Bershadyetal2000,Conselice2003,Grahametal2005}:

\begin{equation}
C=5*\log_{10}(\frac{r_{80\%}}{r_{20\%}}).
\end{equation}

Concentration of stellar light, in rest-frame {\it B}-band, is thus computed for all the galaxies in the sample. The higher this measure, larger is the fraction of total light contained in the central region. 

\subsection{Computing Asymmetry}
Asymmetry in the stellar light of disc galaxies arises from features like bars, star forming clumps, spiral arms, rings etc. It is also observed to be higher for galaxies which are going through interactions or mergers with companion galaxies or accreting satellites (or non-virialized baryonic matter) from the inter galactic medium \citep{Conselice2003,Reichardetal2008,Lotzetal2008}. Since the presence of these features, as well as interactions and mergers, are the expected cause of bulge formation and evolution, tracking the evolution of the asymmetry measure and its relationship with other parameters is of utmost importance.

To compute asymmetry of stellar light in a disc galaxy, we follow the procedure given in \citet{Conseliceetal2000}. We take the cutout galaxy image and rotate it around its centre by 180 degrees. The extraction radius for rotation is given as the Petrosian radius. The rotated image is then subtracted from the main image to get the residual image. The flux from this residual image is a measure of the flux from the asymmetric features of the galaxy. This flux is normalized with respect to the total flux from the main image to get the asymmetry parameter.

For the asymmetry parameter to be meaningfully comparable for the range of disc galaxies, all the sources of probable biases need to be removed \citep{Conselice2003}. The first concern is that the extraction radius should be bias-free, and thus, we choose the Petrosian radius. The second concern is the centre for rotation, which can produce spurious results if not chosen properly. To minimize this effect, the centre for rotation is found in an iterative manner: it is the position for which the asymmetry of the source attains a global minimum.

Another concern is noise, in terms of the background asymmetry. To take that into account, the asymmetry must also be computed from an empty background patch and subtracted from the source's asymmetry. The difficulty with estimating the background asymmetry value is that even if there is a small amount of diffuse light which is more concentrated on one side of the background patch, the asymmetry value increases by a considerable amount. To avoid this issue, we identify relatively clear background patches from a number of galaxy images which are chosen such that they cover a wide range of RA and Dec. The asymmetry is computed on all such patches according to the procedure described above. The mean of these values is then used as the background asymmetry. This procedure is done separately for the galaxy samples in the three redshift ranges.

The asymmetry of the stellar light distribution of the galaxies, in rest-frame {\it B}-band, is thus computed for our sample.

\subsection{Computing effective stellar mass densities}
The stellar mass density of a galaxy is seen to be correlated with its integrated colour, S\'ersic index, concentration, environmental density, as well as star formation rate \citep{Kauffmannetal2003,Brinchmannetal2004,Baldryetal2006,Driveretal2006,Bamfordetal2009}. The stellar mass density inside the effective radius (termed here as effective stellar mass density, ESMD), being highly correlated with the bulge properties of the disc galaxy, is a useful parameter to be examined.

To compute this parameter, the first thing is to obtain the stellar mass. For that, we need to multiply total luminosity in a given band with the corresponding stellar mass-to-light ratio of the galaxy. Total luminosity ($L$) is obtained in units of solar luminosity from the earlier computed absolute magnitudes of the galaxies:
\begin{equation}
M_B=-2.5\log_{10}(\frac{L}{L_\odot})+5.38,
\end{equation}
where $L_\odot$ is the solar luminosity and 5.38 is solar absolute magnitude in rest-frame {\it B}-band. The luminosity inside the effective radius (or half-light radius) is, by definition, half of the total luminosity.

The next step is to obtain the stellar mass-to-light ratio which is known to strongly correlate with the integrated colours \citep{BellanddeJong2001}. To obtain the rest-frame colours, we employ {\it EAZY}, which is a photometric redshift code that provides estimates for rest-frame colour indices \citep{Brammeretal2008}. It compares photometric data to the synthetic photometry of a large range of template spectra and outputs the best match. The important feature of this code is that if the redshift of the source is known with reasonable accuracy, it can be given as a prior or held fixed. Also, the template spectra are based on semi-analytical models and not on empirical spectroscopic samples, which are usually highly biased \citep{Brammeretal2008}. Note that there are other codes similar to EAZY, e.g., FAST \citep{Krieketal2009} and LEPHARE \citep{ArnoutsandIlbert2011}.

The stellar mass-to-light ratio is then computed from the rest-frame $B-V$ colour, using values from \citet{BellanddeJong2001} for the mass-dependent galaxy formation model with bursts \citep{Coleetal2000}. We chose this model because mass dependence is the common feature of the presently acceptable galaxy formation scenarios \citep[][and references therein]{Benson2010}. Also, \citet{BellanddeJong2001} selected this model as their default model claiming that this model reproduces the trends in age and metallicity with respect to surface brightness with the least scatter for local spiral galaxies. 

We, thus, obtain the total stellar mass for each galaxy in our sample using the stellar mass-to-light ratio computed above. This method of obtaining stellar masses is extensively used in extra-galactic studies where well resolved spectral data is not available. The concern is that we are not taking the galaxy colour gradient into account by opting for its global colour. This may lead to an underestimation of mass, however, it should not affect our analysis of relative increase in mass and density. 

The stellar mass inside effective radius is half of the total stellar mass. This mass is then divided by the area within the effective radius to obtain stellar mass density inside the effective radius (or ESMD) in units of solar mass per square kpc.

\subsection{Checking for accuracy and error computation}
The working of the overall procedure/code written to compute Petrosian radii, concentrations and asymmetries of the sources was tested using artificially created images and real images with known parameters. 

The artificial images are created using Galfit's S\'ersic component. First, we fix the S\'ersic index (at n$=$1) and create images with varying half-light radius, i.e. r$_e$$=$40, 50, 60, 62, 65 and 70 pixels. Since the S\'ersic index is fixed, the Petrosian radius is expected to increase with the increase in half-light radius. This is indeed seen as the Petrosian radius for these images is computed to be 98.46($\pm$4.92), 122.74($\pm$6.14), 146.65($\pm$7.33), 151.47($\pm$7.57), 158.44($\pm$7.92) and 169.91($\pm$8.49) pixels respectively. 

Similarly, for fixed half-light radius (at r$_e$$=$50 pixels), we create images with varying S\'ersic index, i.e. n$=$0.8, 1.2, 1.6, 2.0 and 2.4. Here, since the radius containing half of the light is held fixed, the concentration of the galaxy is expected to increase with the increase in its S\'ersic index. This is indeed reported as the concentration index for these images is computed to be 2.65($\pm$0.07), 3.01($\pm$0.08), 3.29($\pm$0.09), 3.52($\pm$0.09) and 3.71($\pm$0.10) respectively.

Next, we create images with different levels of asymmetry. This is done by keeping all the parameters (apparent-magnitude, half-light-radius, S\'ersic index) fixed and adding Fourier modes. A detailed discussion of the representation of various asymmetric galactic features with Fourier modes is provided in \citet{Pengetal2010}. We add the first Fourier mode of varying amplitudes, i.e. 0.07, 0.09, 0.11, 0.13, 0.15, 0.17 and 0.19. The asymmetry measure responds favourably such that it increases with the increasing amount of the Fourier mode amplitude. It is computed to be 0.080, 0.103, 0.125, 0.147, 0.169, 0.191 and 0.213 respectively with error in the range of $\pm$0.045 to 0.048. The procedure, thus, reproduces the measures with reasonable accuracy, responding well to the small shifts produced in the structural properties. 

A concern relating to the accurate computation of these parameters at high redshift is the cosmological surface brightness dimming, which may lead to the non detectability of faint features. However, the surface brightness evolution, reported to be 1-1.5 mag since $z=1$ \citep{Bardenetal2005,Melbourneetal2007,Sachdeva2013} is expected to counter this dimming. Another concern is that of low resolution of galaxies at high redshift. It should not affect our study (till $z\sim1$) because the images are from overlapping 5-orbit depth GOODS survey using HST-ACS, which provides a resolution of 0.03 arcsec per pixel. This is verified in a quantitative way by \citet{Conselice2003}. They simulated the local bright galaxies to higher redshifts as per how these galaxies would be imaged by various surveys. Then they compared the values measured at $z\sim0$ to the values measured at various redshifts. For $z=1$ (for GOODS HST ACS), they report a marginal change of 0.10$\pm$0.18 in the concentration measure and a change of -0.03$\pm$0.07 in the asymmetry measure. Thus, the indices are highly reproducible with negligible scatter.

The error bars in the measurement of these parameters stem from the uncertainties involved in determining the total flux associated with the pixels of interest and also in the selection of the pixels of interest. The error associated with the measurement of flux using the aperture photometry package {\it apphot} of IRAF is calculated using:
\begin{equation}
\begin{split}
error & =\sqrt( {\rm counts}/{\rm gain} + {\rm area}*{\rm stdev}^2\\
& \quad + {\rm area}^2*{\rm stdev}^2/{\rm nsky} ),
\end{split}
\end{equation} 
where gain is in electrons per adu and area (of the aperture) in square pixels, stdev is the standard deviation of the sky counts and nsky is the number of sky pixels.

Out of the total sample of 570 galaxies, the algorithm/procedure did not converge to give the parameter values for 3 galaxies, reporting a floating point zero error. This is an extreme case caused for sources with central point of such high brightness that the inner radius goes to zero.

\subsection{Overall data sample}
\begin{figure}
\mbox{\includegraphics[width=85mm]{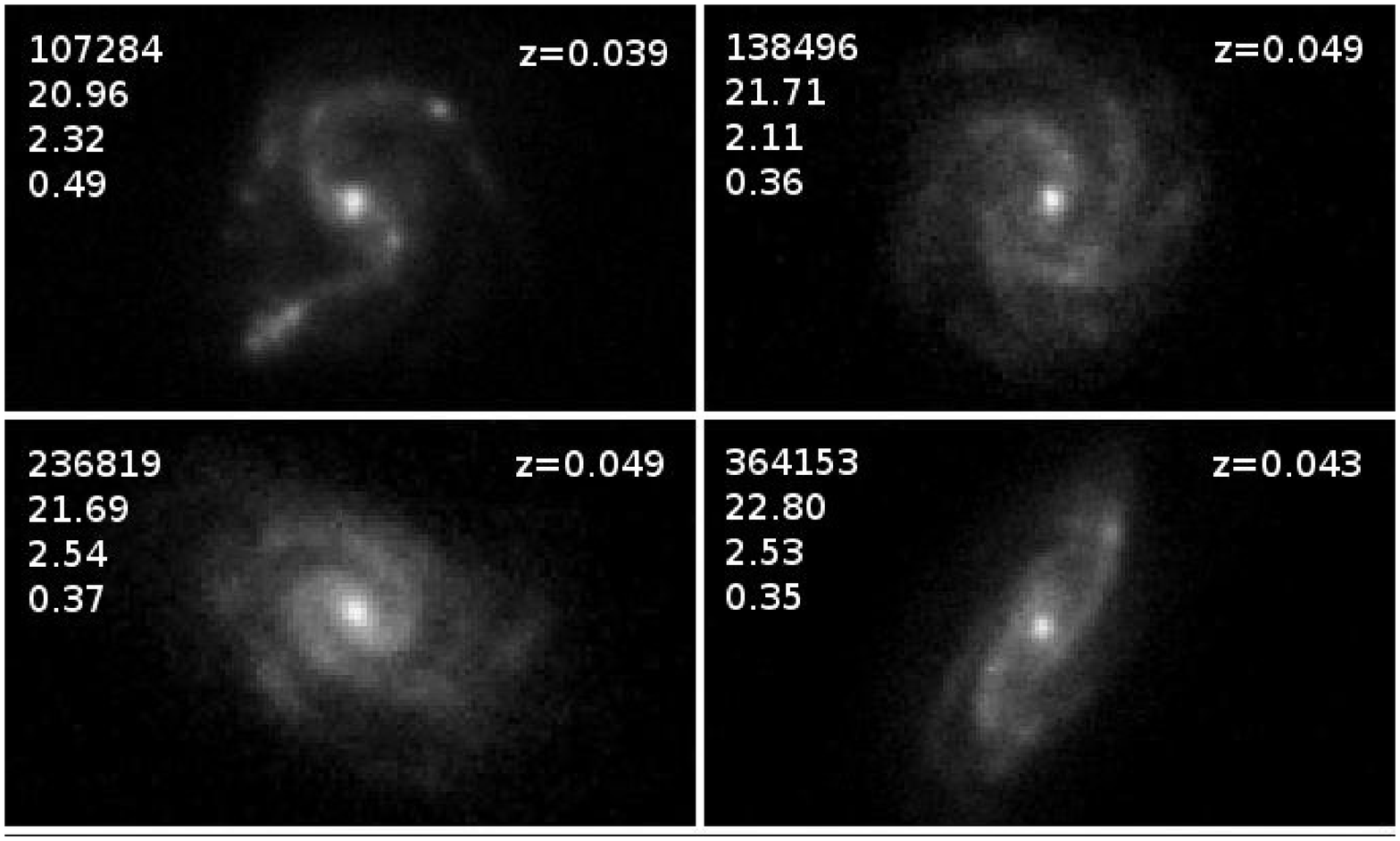}}
\mbox{\includegraphics[width=85mm]{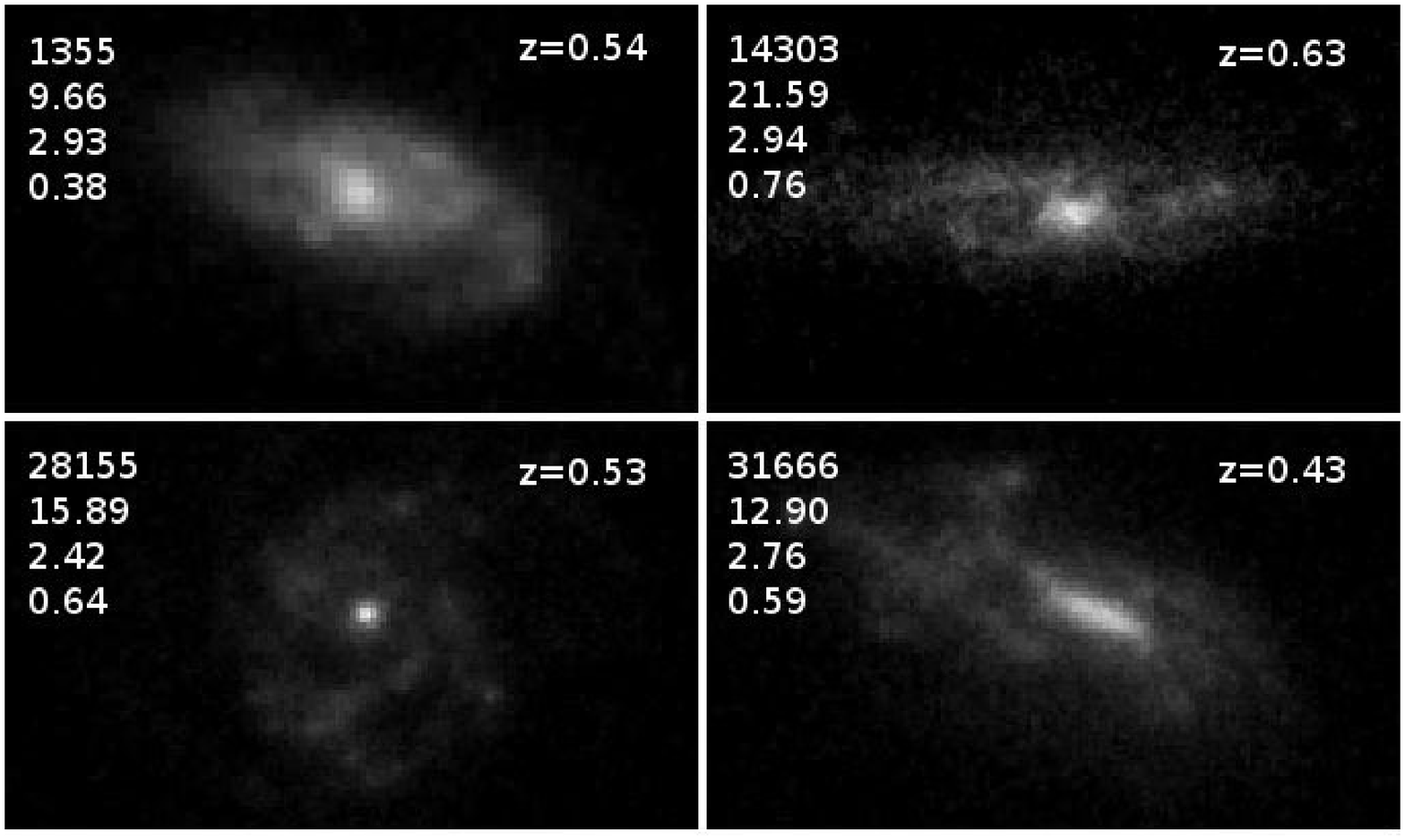}}
\mbox{\includegraphics[width=85mm]{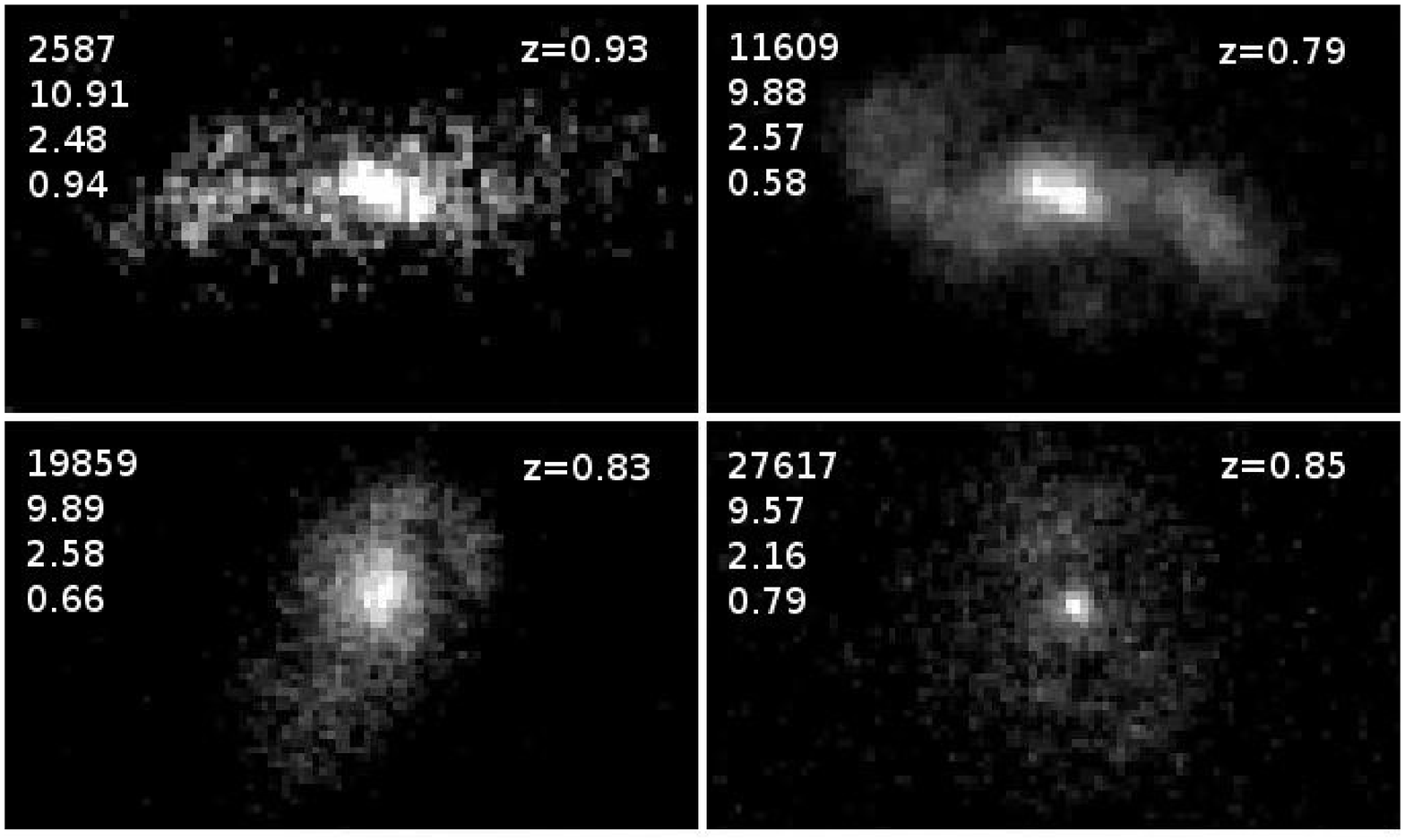}}
\caption{Some of the bulgeless disc galaxies lying in the three redshift ranges (0.02-0.05, 0.4-0.77 and 0.77-1.0) are shown. The non-parametric measures computed for each source are shown at the top left corner. They are in this particular order: ID (as per the NYU-VAGC and GOODS HST-ACS catalogs), Petrosian radius, Concentration and Asymmetry. Precise positional information of the galaxies is provided in Table~\ref{imagepositions}. HST-ACS galaxies cover $\sim$7 arcsec (out of the 10 arcsec image cutout) and SDSS galaxy images cover $\sim$2 arcmin (out of the 3 arcmin image cutout).}
\label{bulgelessimages}
\end{figure}

\begin{figure}
\mbox{\includegraphics[width=85mm]{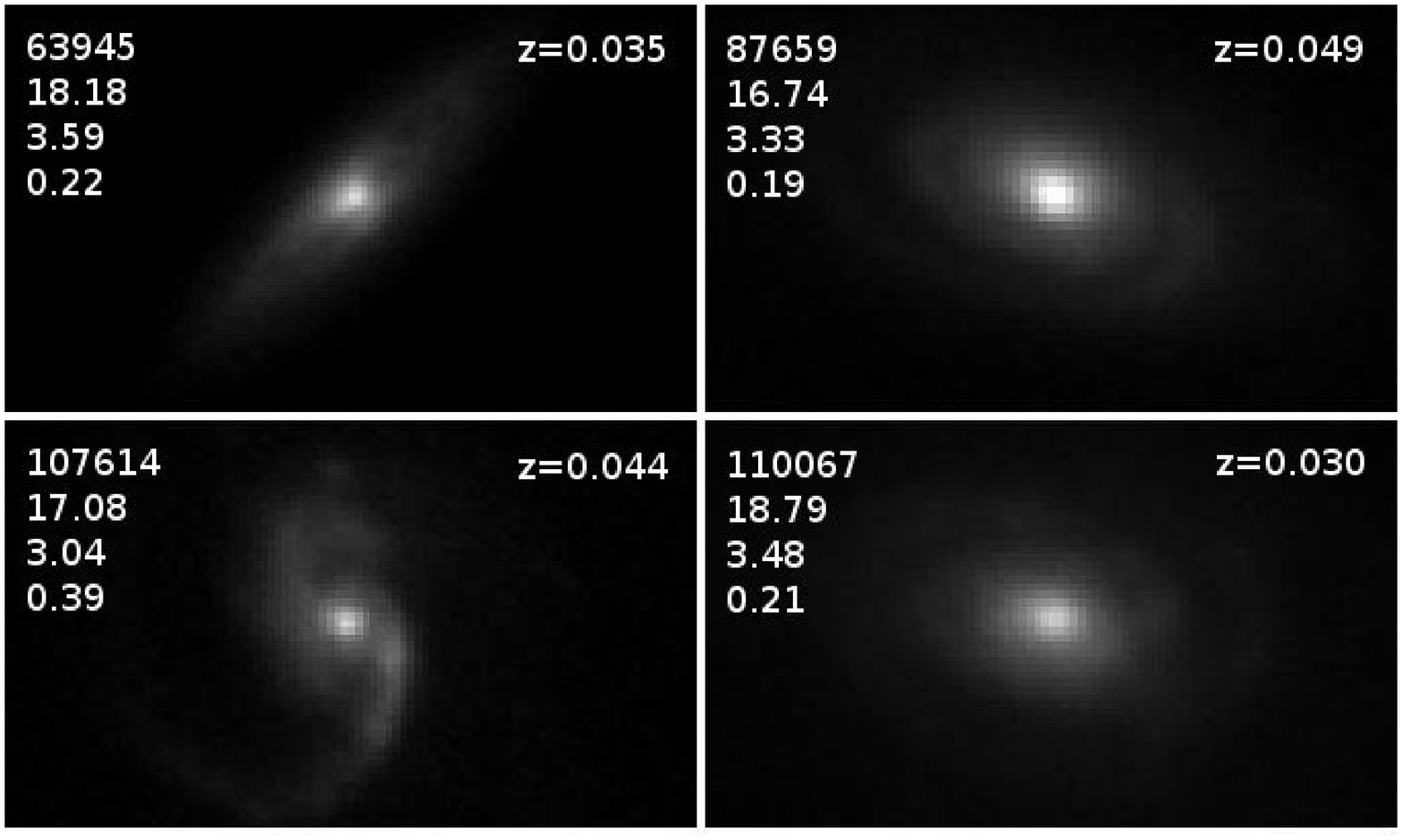}}
\mbox{\includegraphics[width=85mm]{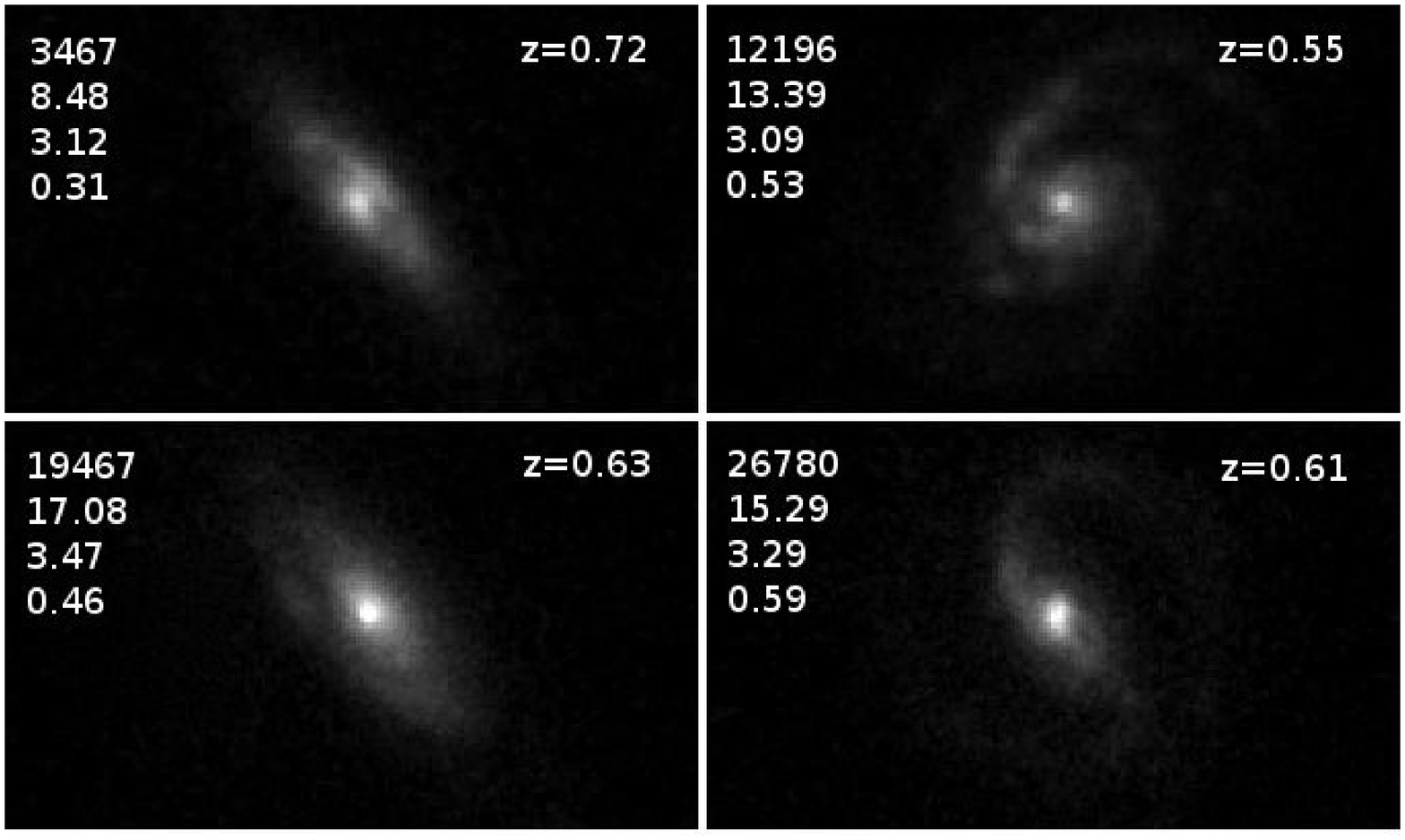}}
\mbox{\includegraphics[width=85mm]{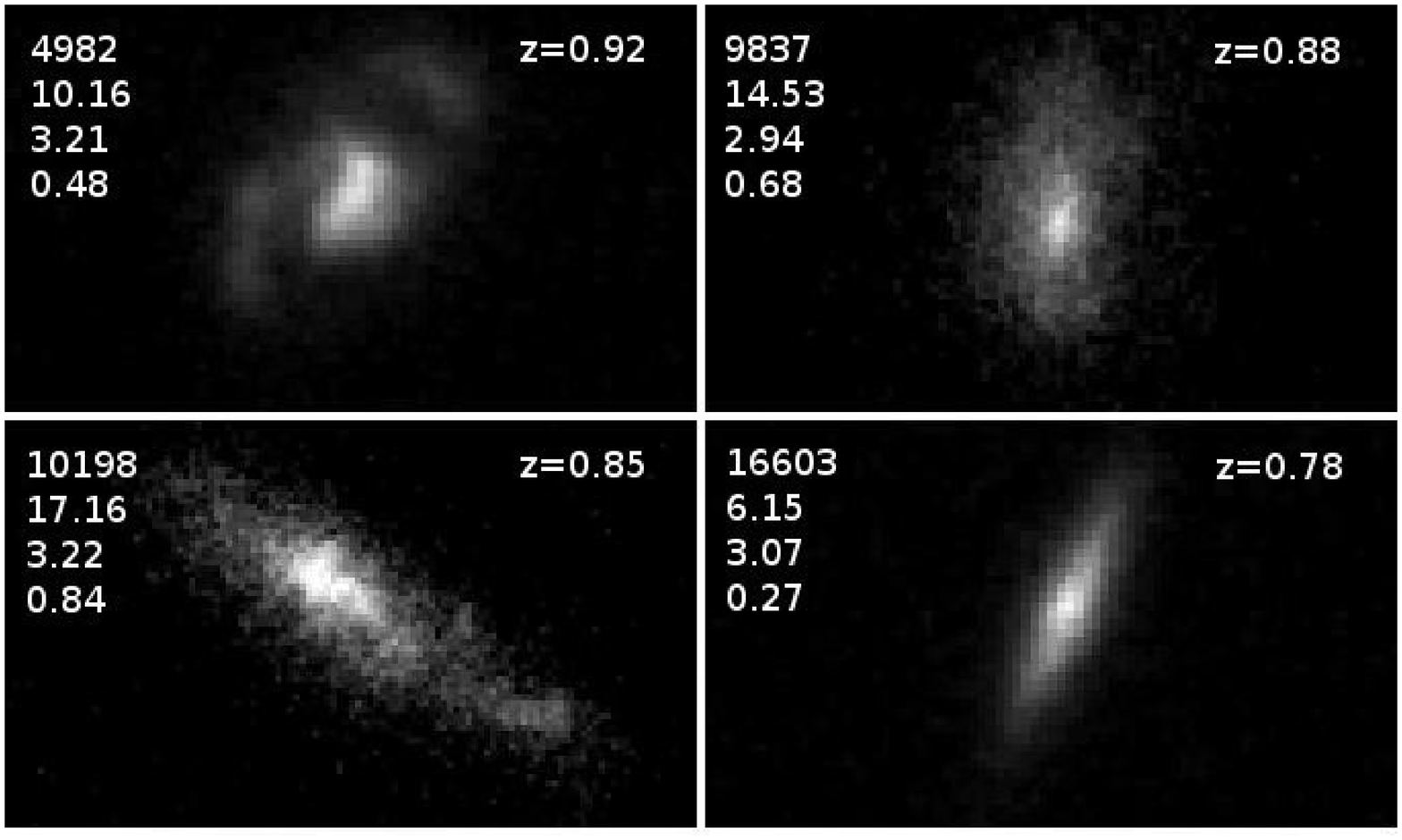}}
\caption{Some of the normal disc galaxies lying in the three redshift ranges (0.02-0.05, 0.4-0.77 and 0.77-1.0) are shown. The non-parametric measures computed for the source are shown at the top left corner. They are in this particular order: ID (as per the NYU-VAGC and GOODS HST-ACS catalogs), Petrosian radius, Concentration and Asymmetry. Precise positional information of the galaxies is provided in Table~\ref{imagepositions}. HST-ACS galaxy images cover $\sim$7 arcsec (out of the 10 arcsec image cutout) and SDSS galaxy images cover $\sim$2 arcmin (out of the 3 arcmin image cutout).}
\label{normalimages}
\end{figure}

We obtained parameters for overall 567 bright (M$_B$$\leq$-20) bulgeless (i.e. without classical bulge) and normal (i.e. with classical bulge) disc dominated galaxies in the three redshift ranges (263 in 0.77$\leq z<$1.0, 203 in 0.4$\leq z<$0.77, 101 in 0.02$\leq z<$0.05) in rest-frame {\it B}-band. We have their redshifts, absolute magnitudes, half-light radii in kpc and S\'ersic index from \citet{Sachdeva2013}. We have now computed their Petrosian radii in kpc, concentration, asymmetry, total stellar mass and effective stellar mass density. Some of the bulgeless and normal disc galaxy images from the sample are shown in Fig.~\ref{bulgelessimages} and Fig.~\ref{normalimages} for the three redshift ranges. The computed parameters along with the associated errors for these images are provided in Table~\ref{imagepositions}. In the next section, we present the results that provide insights into bulge formation and evolution occurring in disc galaxies, by examining the evolution and relationships of these parameters.

\begin{table*}
\centering
\begin{minipage}{200mm}
\caption{Parameters for galaxies shown in Fig.~\ref{bulgelessimages} and Fig.~\ref{normalimages}.}
\begin{tabular}{@{}lllllllllll@{}}
\hline
Source & ID & RA (J2000) & Dec (J2000) & $z$ & $C$ & $C-err$ & $A$ & $A-err$ & $R_p$ & $R_p-err$\\
 & & degrees & degrees & & & & & & Kpc & Kpc\\
\hline
SDSS & 107284 & 18.182632 & 15.70782 & 0.039 & 2.318 & 0.118 & 0.492 & 0.026 & 20.962 & 1.048\\
SDSS & 138496 & 140.67966 & 57.51786 & 0.049 & 2.106 & 0.128 & 0.359 & 0.036 & 21.709 & 1.085\\
SDSS & 236819 & 318.44444 & -5.8170589 & 0.049 & 2.543 & 0.157 & 0.367 & 0.034 & 21.686 & 1.084\\
SDSS & 364153 & 148.98351 & 53.693652 & 0.044 & 2.531 & 0.127 & 0.356 & 0.027 & 22.805 & 1.140\\
SDSS & 63945 & 226.72858 & 0.1863334 & 0.035 & 3.586 & 0.182 & 0.222 & 0.016 & 18.186 & 0.909\\
SDSS & 87659 & 231.89702 & 0.28918961 & 0.049 & 3.333 & 0.254 & 0.192 & 0.009 & 16.739 & 0.837\\
SDSS & 107614 & 28.966593 & 14.940272 & 0.044 & 3.042 & 0.203 & 0.397 & 0.018 & 17.078 & 0.854\\
SDSS & 110067 & 42.326159 & -8.1749502 & 0.030 & 3.485 & 0.156 & 0.211 & 0.008 & 18.788 & 0.939\\
HST-ACS & 1355 & 53.0167785 & -27.7189942 & 0.538 & 2.93 & 0.192 & 0.376 & 0.034 & 9.661 & 0.483\\
HST-ACS & 14303 & 53.1091768 & -27.8529078 & 0.629 & 2.937 & 0.096 & 0.759 & 0.119 & 21.586 & 1.079\\
HST-ACS & 28155 & 53.190166 & -27.7349438 & 0.530 & 2.417 & 0.104 & 0.643 & 0.077 & 15.897 & 0.795\\
HST-ACS & 31666 & 53.2226372 & -27.8475767 & 0.432 & 2.755 & 0.121 & 0.598 & 0.058 & 12.901 & 0.645\\
HST-ACS & 2587 & 53.0311719 & -27.7357189 & 0.935 & 2.483 & 0.209 & 0.937 & 0.197 & 10.912 & 0.546\\
HST-ACS & 11609 & 53.093944 & -27.8727229 & 0.789 & 2.573 & 0.202 & 0.581 & 0.074 & 9.875 & 0.494\\
HST-ACS & 19859 & 53.1384525 & -27.6806527 & 0.828 & 2.581 & 0.207 & 0.657 & 0.111 & 9.887 & 0.494\\
HST-ACS & 27617 & 53.1859173 & -27.7756062 & 0.848 & 2.155 & 0.186 & 0.791 & 0.16 & 9.571 & 0.478\\
HST-ACS & 3467 & 53.0391609 & -27.7100294 & 0.716 & 3.12 & 0.276 & 0.312 & 0.036 & 8.484 & 0.424\\
HST-ACS & 12196 & 53.0975165 & -27.7212663 & 0.551 & 3.09 & 0.156 & 0.531 & 0.048 & 13.391 & 0.669\\
HST-ACS & 19467 & 53.1361845 & -27.836454 & 0.633 & 3.466 & 0.139 & 0.46 & 0.05 & 17.084 & 0.854\\
HST-ACS & 26780 & 53.179339 & -27.9235978 & 0.615 & 3.296 & 0.151 & 0.596 & 0.07 & 15.290 & 0.764\\
HST-ACS & 4982 & 53.050917 & -27.7724075 & 0.924 & 3.211 & 0.246 & 0.482 & 0.043 & 10.156 & 0.508\\
HST-ACS & 9837 & 53.0831774 & -27.7471694 & 0.883 & 2.937 & 0.163 & 0.678 & 0.108 & 14.526 & 0.726\\
HST-ACS & 10198 & 53.0854551 & -27.6830331 & 0.847 & 3.224 & 0.138 & 0.837 & 0.129 & 17.162 & 0.858\\
HST-ACS & 16603 & 53.120832 & -27.8230569 & 0.783 & 3.067 & 0.367 & 0.272 & 0.033 & 6.149 & 0.307\\
\hline
\label{imagepositions}
\end{tabular}
\end{minipage}
\end{table*}


\section{Results}
The formation and evolution of bulges in disc galaxies can be probed by examining the mutual evolution and relationship of those parameters whose measure is associated with bulge properties. We, thus, examine the evolution of size, concentration, inner stellar density and asymmetry.
\subsection{Size evolution}
The sizes of the massive disc galaxies in the universe are seen to undergo a dramatic increase with time \citep[e.g.][]{Trujilloetal2007,Buitragoetal2008,VanDokkumetal2010,Carrascoetal2010,Cassataetal2013}. Some studies have attempted to explain this increase through inside-out processes, major mergers, minor mergers, accretion, AGN processes etc. \citep{Hopkinsetal2009,Kavirajetal2009,Blucketal2012,Ownsworthetal2012,McLureetal2013}. 

The mechanisms that lead to the overall increase in disc length are also expected to produce variations in the inner region. For example, secular evolution driven by various asymmetric structures causes galactic discs to expand on the outside and contract on the inside \citep{Tremaine1989,VanDokkumetal2010,Kormendy2013}. To understand the relative role of the mechanisms in disc growth, the increase in the total optical extent should be tracked with the change in the inner region properties.

We, therefore, examine the increase in the Petrosian radius of the galaxy relative to the increase in its half-light radius. This relative increase will be seen separately for discs with and without classical bulges at the three redshift ranges (0.02-0.05, 0.4-0.77, 0.77-1.0).

\begin{figure*}
\mbox{\includegraphics[width=60mm]{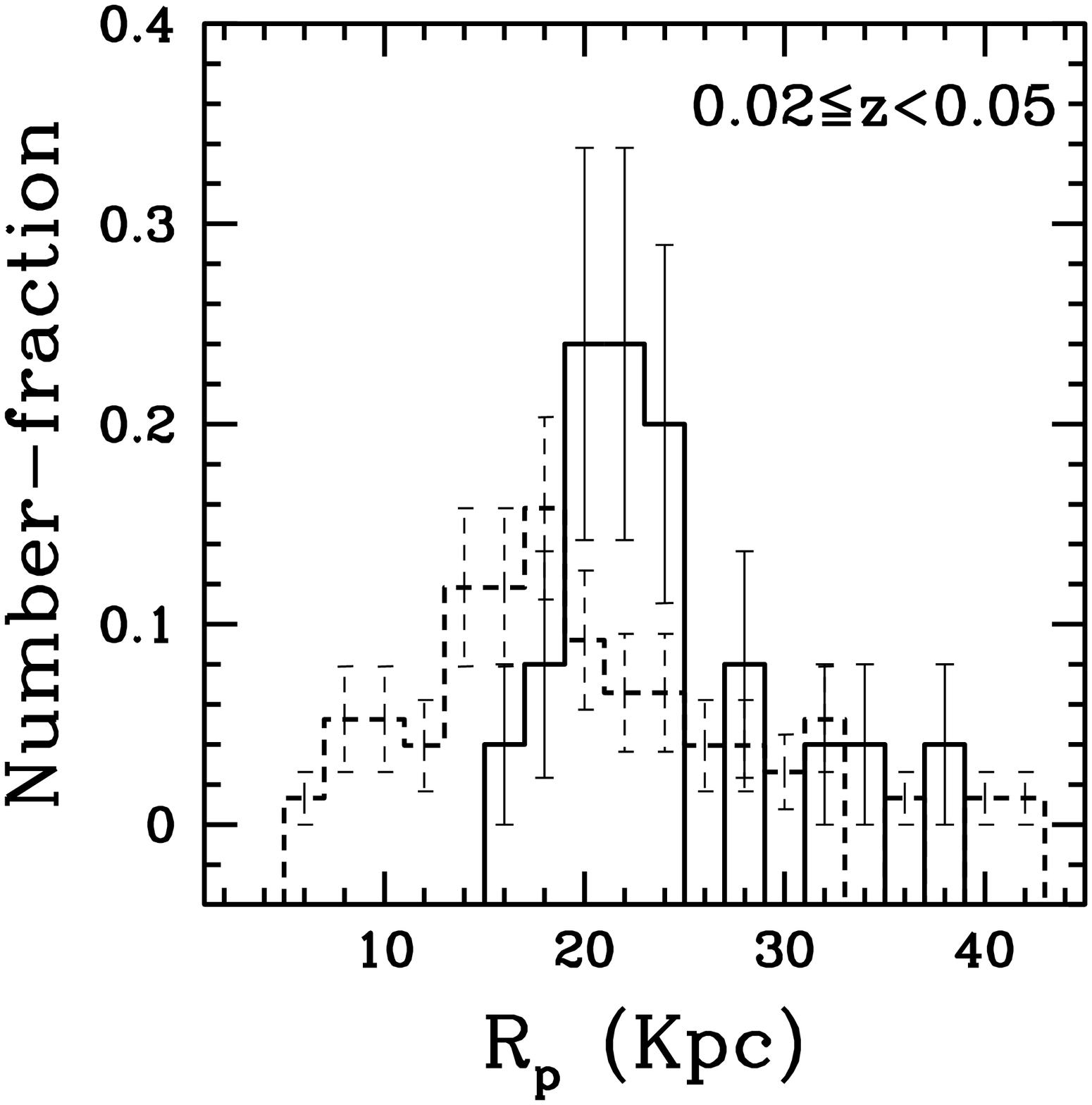}}
\mbox{\includegraphics[width=60mm]{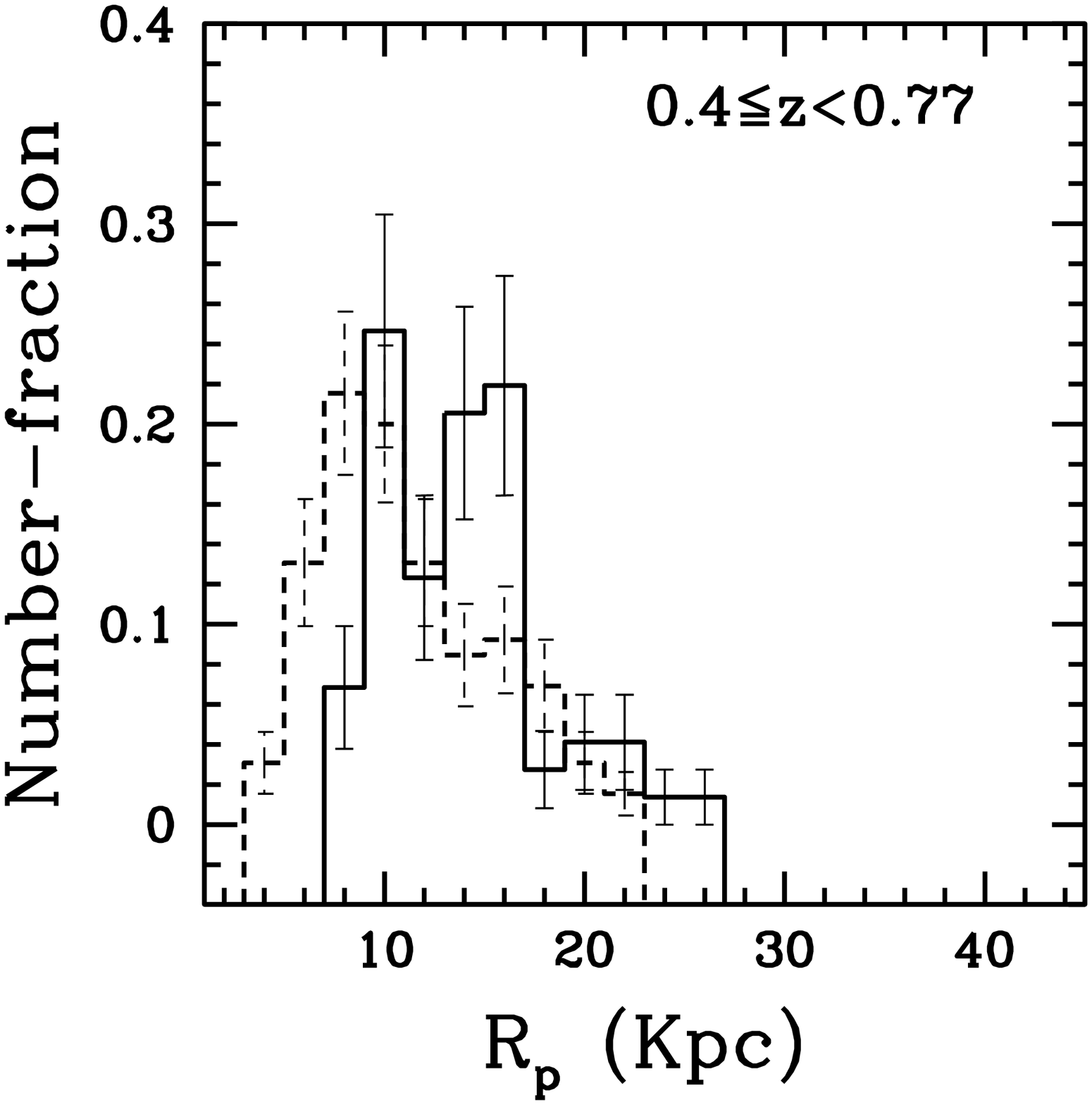}}
\mbox{\includegraphics[width=60mm]{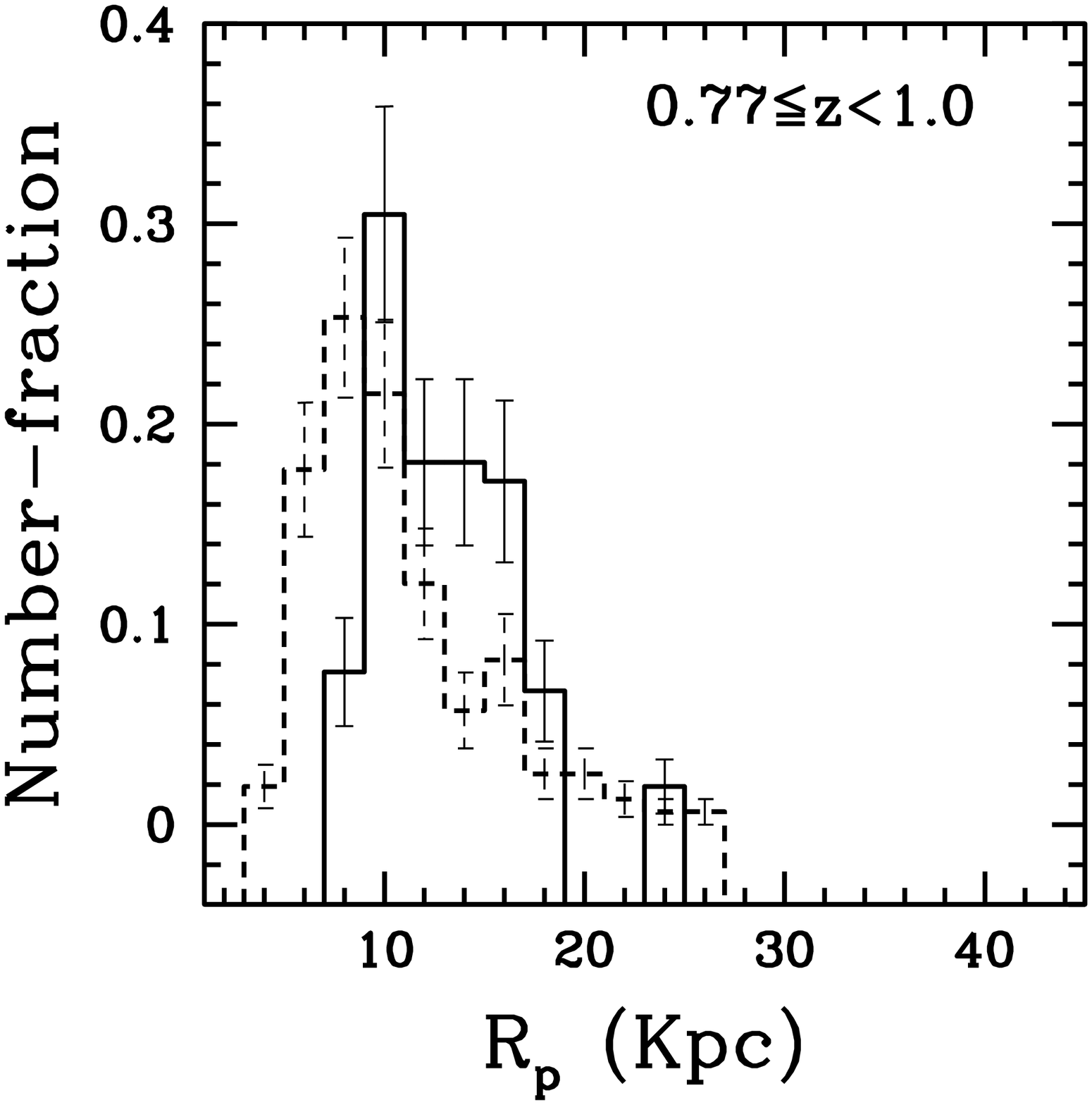}}
\mbox{\includegraphics[width=60mm]{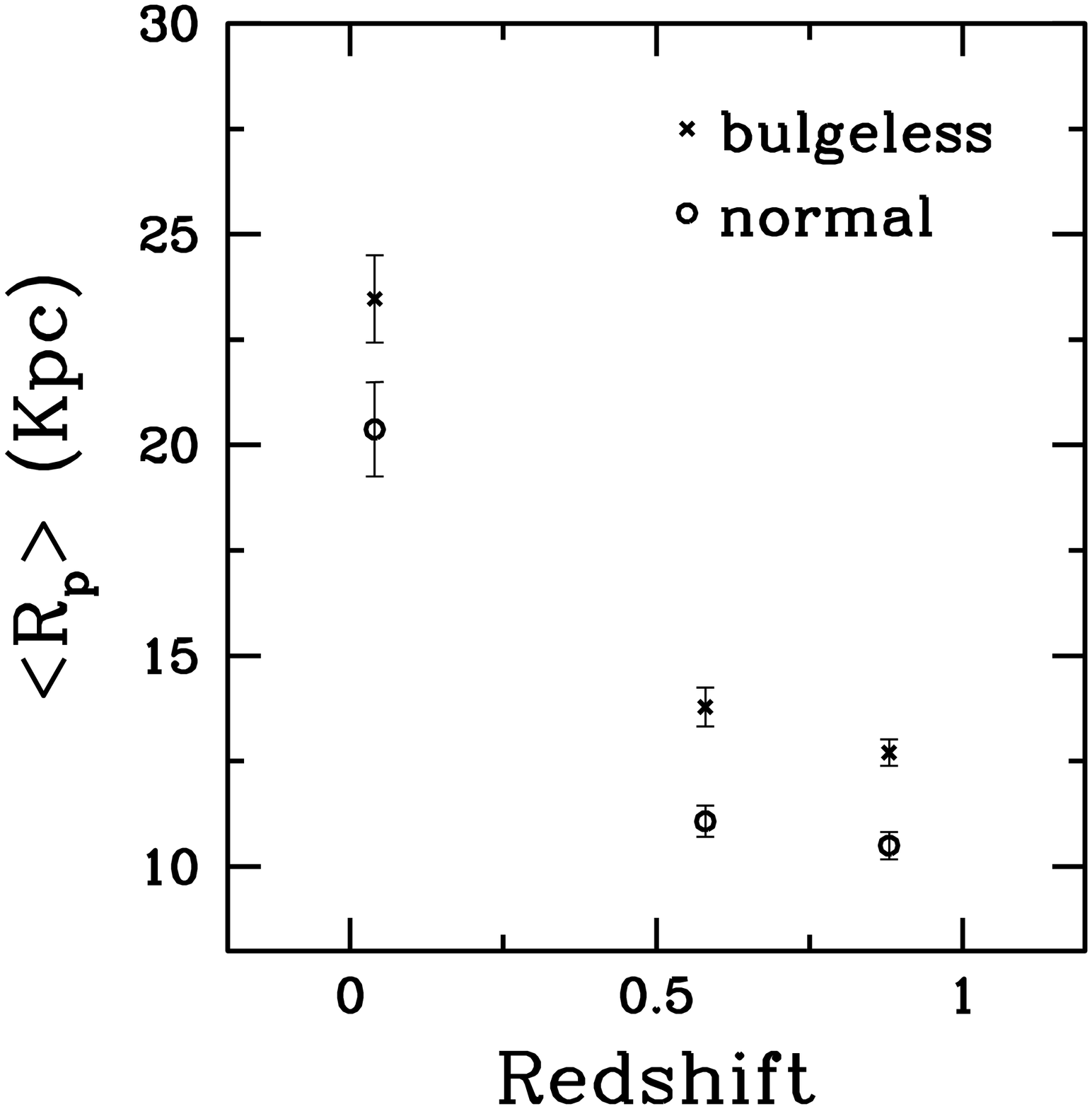}}
\caption{Distributions of Petrosian radius for bulgeless disc (solid lines) and normal disc (dashed lines) galaxies for the three redshift ranges. The distribution of the means of the two samples with redshift is also shown. There is a significant increase in sizes for both samples with time.}
\label{histpetr}
\end{figure*}

The distribution of the Petrosian radius for bulgeless and normal disc galaxies is shown in Fig.~\ref{histpetr} for the three redshift ranges. For the bulgeless disc galaxies it increases by 85($\pm$4)\% (from 12.71 kpc to 23.46 kpc) from $z\sim0.9$ to the present epoch. Over the same time range, for the normal disc sample, it increases by 94($\pm$6)\% (from 10.50 kpc to 20.37 kpc) (see Table~\ref{meanpetrre}).

Thus, both morphological types show a significant increase in size. However, the evolution is faster and leads to a more effective increase in size for the normal disc galaxies. Nevertheless, it is interesting to note that bulgeless disc galaxies are larger than the normal disc galaxies, on average, at all redshift bins.

In contrast to the Petrosian radius, the half-light radius for both bulgeless and normal disc samples shows a minor increase of 28($\pm$2)\% (from 5.43 kpc to 6.97 kpc) and 22($\pm$1)\% (from 3.72 kpc to 4.55 kpc), respectively, from $z\sim0.9$ to the present epoch (Table~\ref{meanpetrre}).

If we see the relative quantity, i.e. the ratio of Petrosian radius to half-light radius, it increases from 2.34 to 3.36 from $z\sim0.9$ to the present epoch for bulgeless disc galaxies. However, the increase is larger, from 2.82 to 4.48, for the normal disc galaxies. For these galaxies, their total radius becomes $\sim$4.5 times their half-light radius, as we reach the present epoch.

\begin{figure*}
\mbox{\includegraphics[width=55mm]{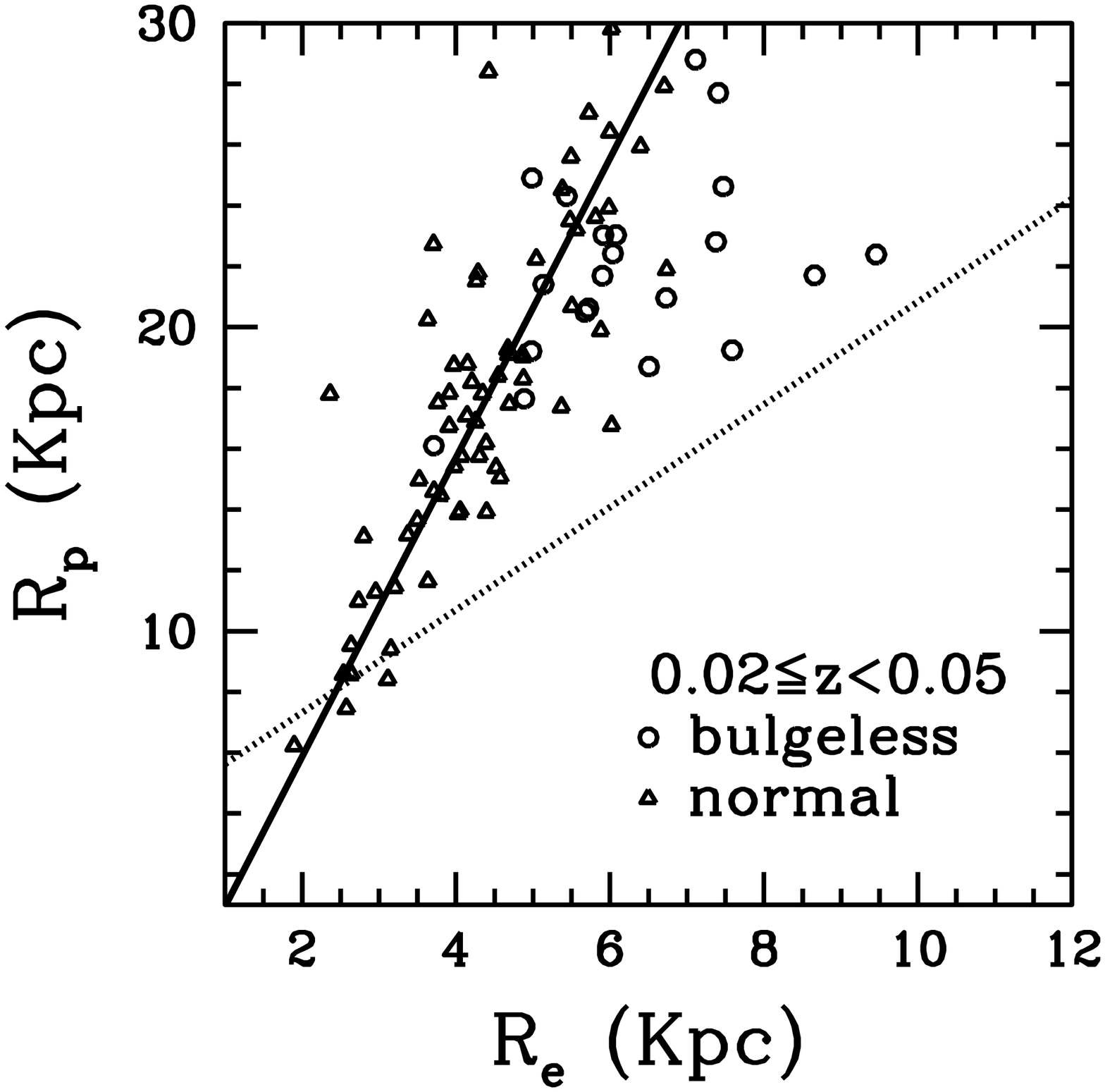}}
\mbox{\includegraphics[width=55mm]{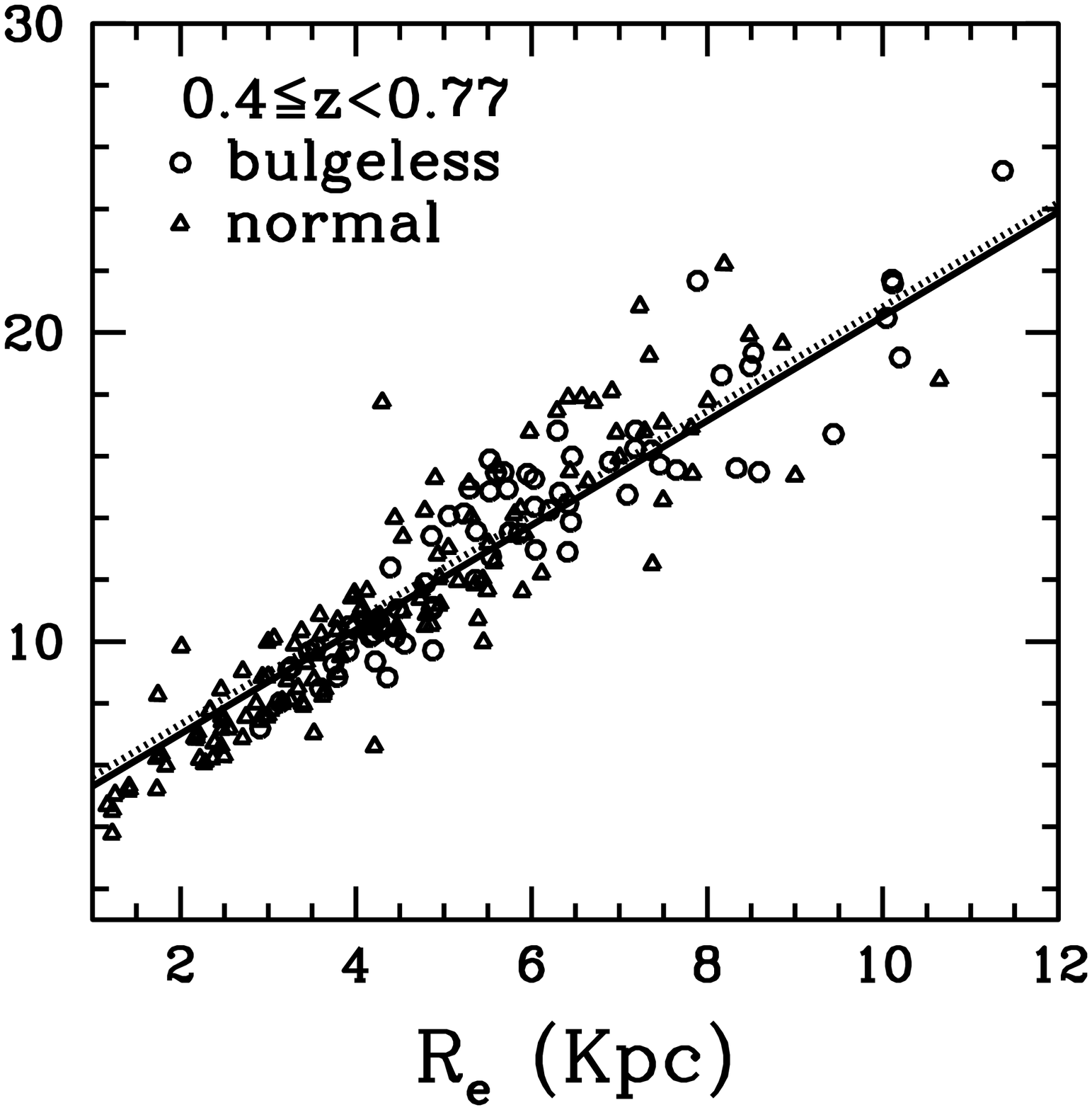}}
\mbox{\includegraphics[width=55mm]{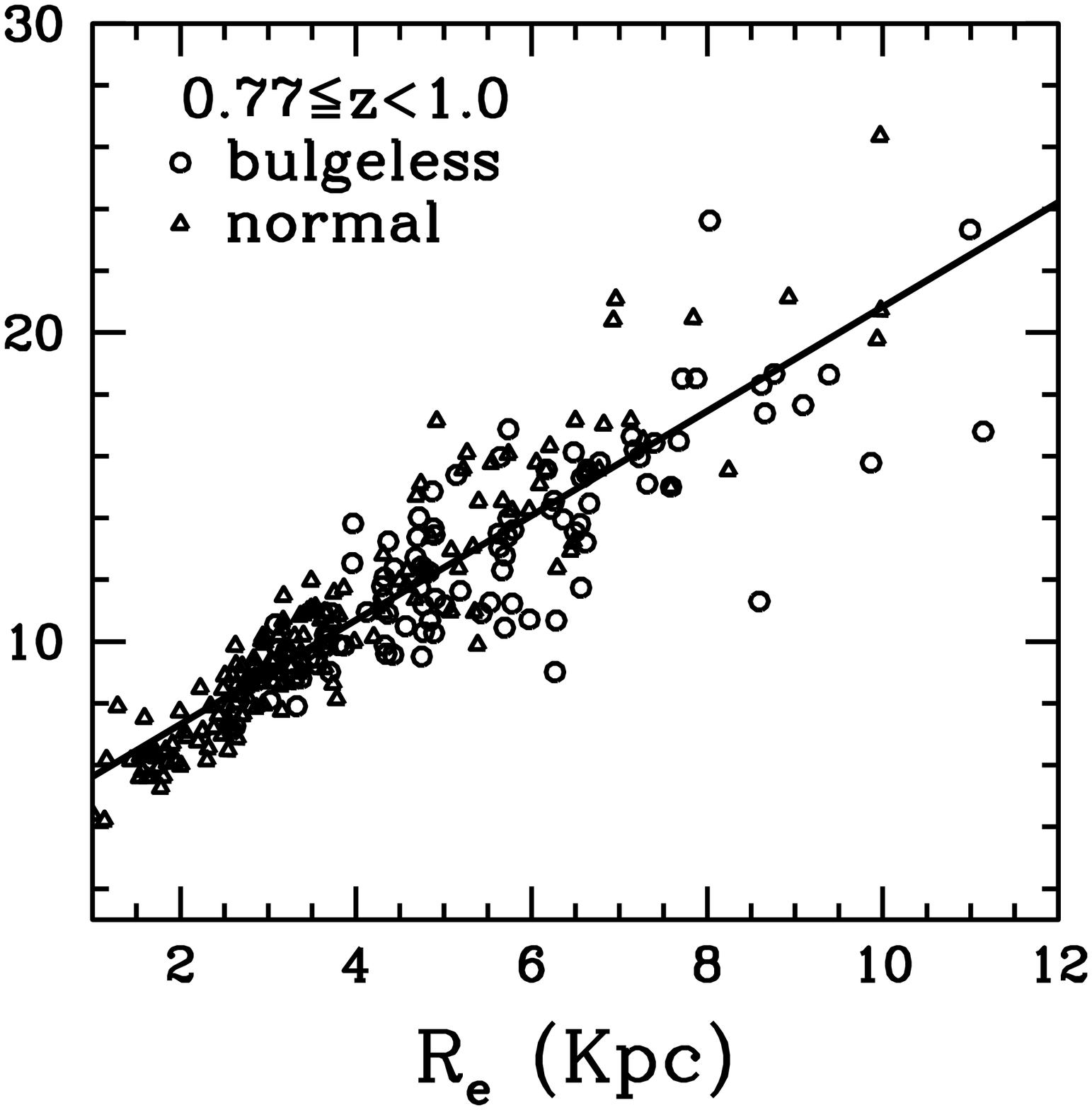}}
\caption{Petrosian radius is plotted against half-light radius for bulgeless and normal disc galaxies in the three redshift ranges. The solid line in each graph marks the linear relation followed by the full sample at that redshift range. The highest redshift range (0.77-1.0) relation is shown on the two lower redshift range (0.4-0.77, 0.02-0.05) plots with dotted lines. It is seen to almost overlap the intermediate redshift range (0.4-0.77) relation. However, for the local sample (0.02-0.05), the slope is entirely different. There the half-light radius is seen to be a much smaller fraction of the Petrosian radius. For the Petrosian-radius, the error scales with the value and is seen to be in the range of $\pm$5\%. For half light radius, the typical error is of $\pm$0.2 kpc.}
\label{petrvsre}
\end{figure*}

The two sizes (half-light radius and Petrosian radius), though computed in totally different ways (parametrically and non-parametrically), are found to be highly correlated in our galaxy sample, as seen in Fig.~\ref{petrvsre}. The linear fit relation found for the highest redshift range (0.77-1.0) almost overlaps the relation found for the intermediate redshift range (0.4-0.77). However, the slope changes drastically for the local redshift range (0.02-0.05): 
\begin{equation}
R_P=1.69(\pm0.06)*R_e+3.62(\pm0.31); 0.4\leq z<1.0,
\end{equation}
\begin{equation}
R_P=4.92(\pm0.68)*R_e-2.04(\pm0.59); 0.02\leq z<0.05.
\end{equation}
The change of slope further signifies that the half-light radius reduces to a much smaller fraction of the Petrosian radius as we reach the present epoch.

Overall, we find that while the total extent of our discs almost doubles with time, the radius containing half the stellar light increases marginally in comparison. This suggests significant peripheral increase, which is seen to be somewhat more pronounced for the normal disc sample.

\begin{table*}
\centering
\begin{minipage}{200mm}
\caption{Mean and median values of Petrosian radius and half-light radius.}
\begin{tabular}{@{}llllllllll@{}}
\hline
Redshift & Disc & No. of & Mean of & Std.dev. & Median & Mean of & Std.dev. & Median\\
range & type & sources & Petrosian-radius & of $R_p$ & of $R_p$ & half-light radius & of $R_e$ & of $R_e$\\
 & & & $<R_p>$(Kpc) & $\sigma$ & & $<Re>$(Kpc) & $\sigma$ &\\
\hline
0.77-1.0 & bulgeless & 105 & 12.706($\pm$0.317) & 3.246 & 12.249($\pm$0.397) & 5.429($\pm$0.179) & 1.844 & 4.902($\pm$0.224)\\
0.77-1.0 & normal & 158 & 10.498($\pm$0.325) & 4.087 & 9.411($\pm$0.407) & 3.725($\pm$0.149) & 1.885 & 3.170($\pm$0.187)\\
0.4-0.77 & bulgeless & 73 & 13.787($\pm$0.459) & 3.922 & 13.871($\pm$0.575) & 5.947($\pm$0.251) & 2.148 & 5.539($\pm$0.314)\\
0.4-0.77 & normal & 130 & 11.075($\pm$0.368) & 4.194 & 10.507($\pm$0.461) & 4.446($\pm$0.193) & 2.208 & 4.080($\pm$0.242)\\
0.02-0.05 & bulgeless & 25 & 23.464($\pm$1.035) & 5.177 & 22.395($\pm$1.297) & 6.972($\pm$0.500) & 2.504 & 6.508($\pm$0.626)\\
0.02-0.05 & normal & 76 & 20.373($\pm$1.119) & 9.761 & 18.244($\pm$1.402) & 4.552($\pm$0.147) & 1.280 & 4.368($\pm$0.184)\\
\hline
\label{meanpetrre}
\end{tabular}
\end{minipage}
\end{table*}

\subsection{Concentration evolution}
The S\'ersic index is extensively used not just for morphological characterization of galaxies but also to study their structural evolution \citep{Shenetal2003,Blantonetal2005a,FisherandDrory2008,Conseliceetal2011,Bruceetal2012,Buitragoetal2013}. Being related to the steepness of the intensity profile, it is a measure of the prominence of the bulge component \citep{Sersic1968,Pengetal2002}. The concentration parameter, however, is seen to be a better estimator of the bulge presence for low redshift galaxies \citep{Conselice2003,Grahametal2005,Gadotti2009}. 

Our full (disc dominated) sample is separated on the basis of S\'ersic index (along with the Kormendy relation). We now examine the evolution in their concentration value. The distribution of the concentration for bulgeless and normal disc galaxies is studied for the three redshift ranges (0.02-0.05, 0.4-0.77, 0.77-1.0). The mean concentration of the bulgeless disc galaxies shows a statistically insignificant increase from $z\sim0.9$ to the present epoch (Table~\ref{meanconcasym}). However, for the normal disc sample it increases by 12.3($\pm$0.3)\% (from 2.77 to 3.11), over the same time range (Table~\ref{meanconcasym}).

Thus, while the average bulgeless disc galaxy concentration remains almost similar over the three redshift ranges, there is a significant (albeit admittedly small) increase in average normal disc galaxy concentration.

\begin{figure*}
\mbox{\includegraphics[width=55mm]{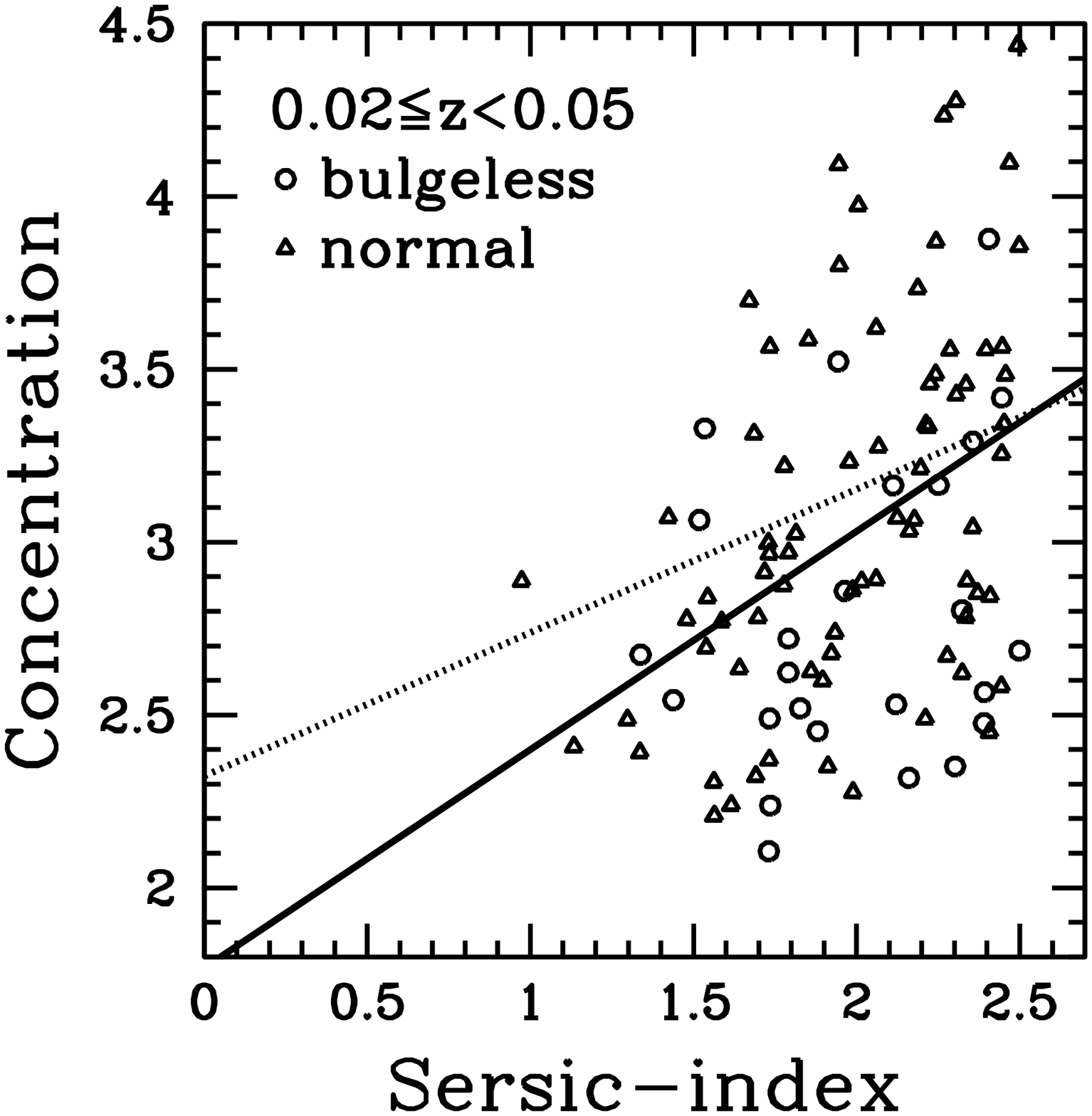}}
\mbox{\includegraphics[width=55mm]{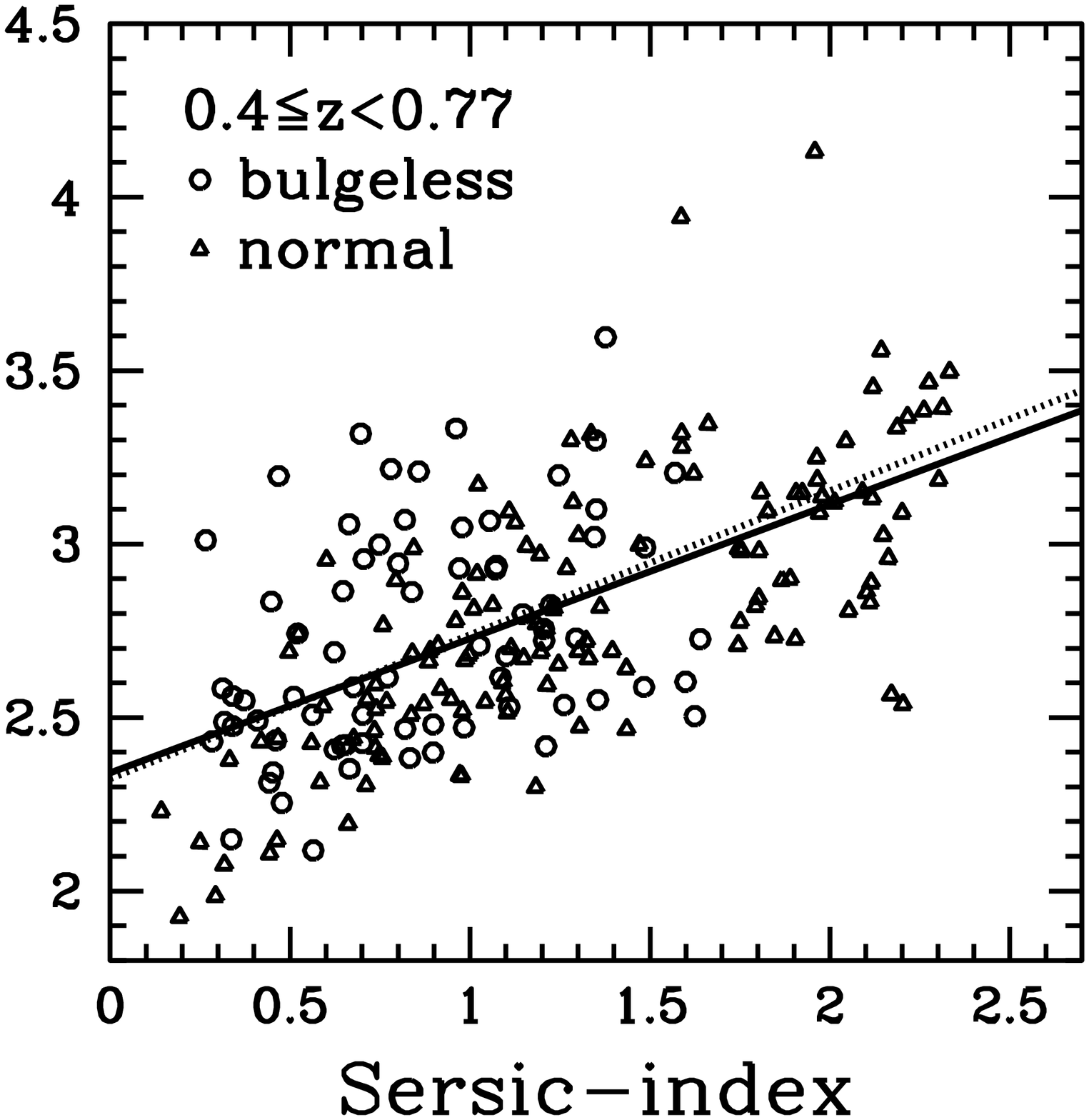}}
\mbox{\includegraphics[width=55mm]{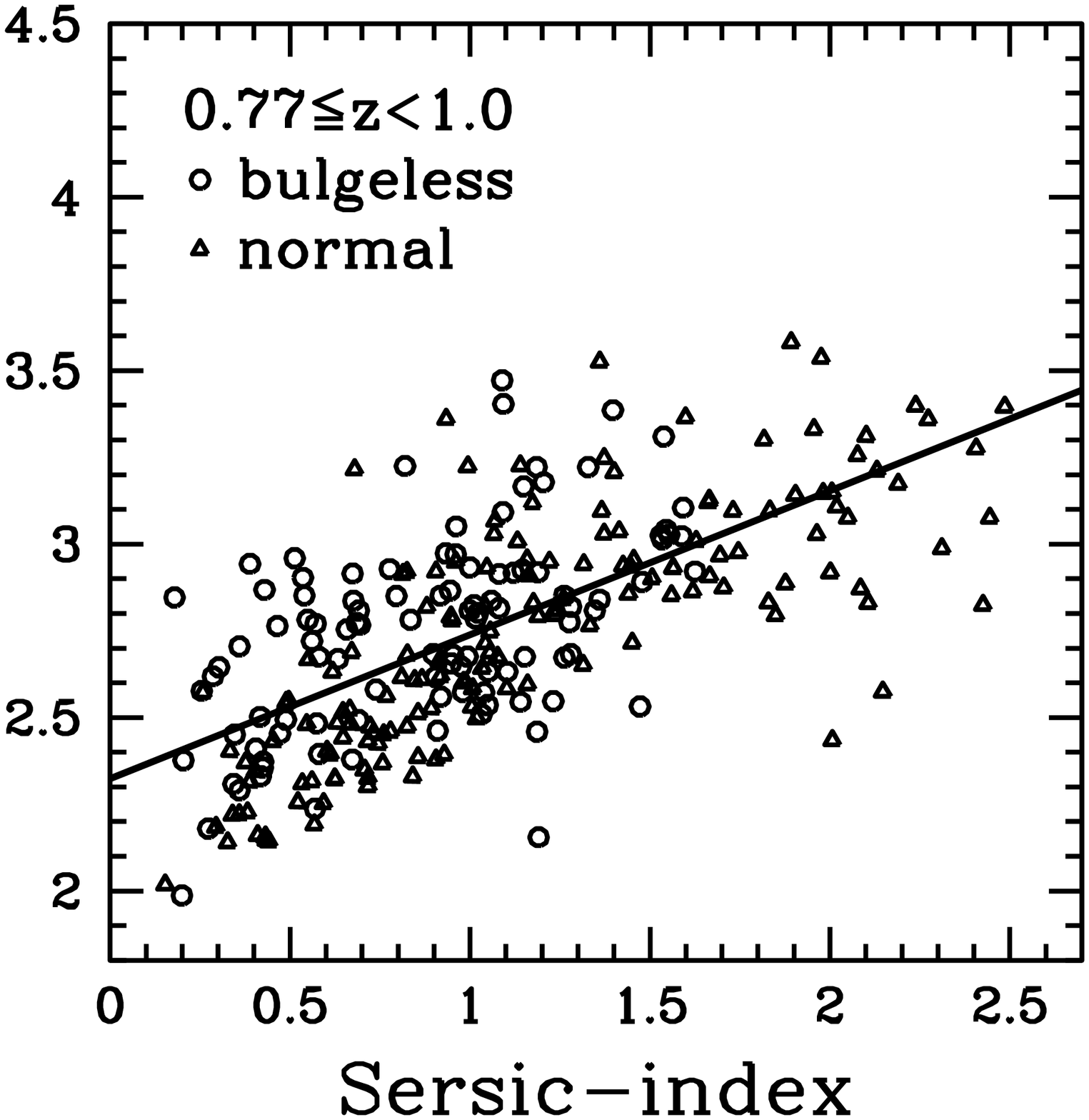}}
\caption{Concentration is plotted against S\'ersic index, n, for bulgeless and normal disc galaxies in the three redshift ranges. The solid line in each graph marks the linear relation followed by the full sample at that redshift range. The highest redshift range (0.77-1.0) relation is shown on the two lower redshift range (0.4-0.77, 0.02-0.05) plots with dotted lines. The two parameters are seen to be highly correlated and the relation almost overlaps for the two high redshift ranges (0.77-1.0, 0.4-0.77). However, the local sample shows a lack of correlation in the two values. For concentration, typical error on the value is $\pm$0.17 and for S\'ersic index it is $\pm$0.05.}
\label{concvsn}
\end{figure*}

While the S\'ersic index has been obtained parametrically, i.e., by fitting a function to the surface brightness profile, the concentration is based on total count ratios, i.e. non-parametrically. We examine the relationship of the two parameters for the full sample (Fig.~\ref{concvsn}) in the three redshift ranges. The two quantities are seen to be well correlated for both morphological types and follow a single relation for both the higher (0.77-1.0) and intermediate (0.4-0.77) redshift ranges:
\begin{equation}
C=0.38(\pm0.03)*n+2.32(\pm0.04), 0.4\leq z<1.0
\end{equation}
The two quantities appear to provide a similar estimate of the bulge component. Thus, according to our study they are equally well deserving to be chosen at high redshifts for morphological determination.

However, the correlation is absent for the local redshift-range (0.02-0.05). This lack of correlation at local redshifts is also reported by Gadotti (2009). We speculate that for highly resolved local galaxies the intensity gradient between the bulge and the disc is enhanced, leading to poorer fits when only a single function is used to fit the entire galaxy.

\begin{table*}
\centering
\begin{minipage}{200mm}
\caption{Mean and median values of concentration and asymmetry.}
\begin{tabular}{@{}llllllllll@{}}
\hline
Redshift & Disc & No. of & Mean of & Std.dev. & Median & Mean of & Std.dev. & Median\\
range & type & sources & concentration & of $C$ & of $C$ & asymmetry & of $A$ & of $A$\\
 & & & $<C>$ & $\sigma$ & & $<A>$ & $\sigma$ &\\
\hline
0.77-1.0 & bulgeless & 105 & 2.748($\pm$0.027) & 0.278 & 2.772($\pm$0.034) & 0.827($\pm$0.014) & 0.145 & 0.831($\pm$0.017)\\
0.77-1.0 & normal & 158 & 2.767($\pm$0.030) & 0.382 & 2.784($\pm$0.037) & 0.568($\pm$0.014) & 0.181 & 0.559($\pm$0.017)\\
0.4-0.77 & bulgeless & 73 & 2.726($\pm$0.037) & 0.317 & 2.677($\pm$0.046) & 0.719($\pm$0.019) & 0.159 & 0.718($\pm$0.024)\\
0.4-0.77 & normal & 130 & 2.819($\pm$0.034) & 0.391 & 2.776($\pm$0.042) & 0.515($\pm$0.016) & 0.188 & 0.494($\pm$0.020)\\
0.02-0.05 & bulgeless & 25 & 2.791($\pm$0.067) & 0.448 & 2.674($\pm$0.084) & 0.412($\pm$0.023) & 0.115 & 0.394($\pm$0.029)\\
0.02-0.05 & normal & 76 & 3.107($\pm$0.064) & 0.560 & 3.010($\pm$0.080) & 0.308($\pm$0.015) & 0.131 & 0.280($\pm$0.019)\\
\hline
\label{meanconcasym}
\end{tabular}
\end{minipage}
\end{table*}

\subsection{Stellar density evolution}
\begin{figure*}
\mbox{\includegraphics[width=60mm]{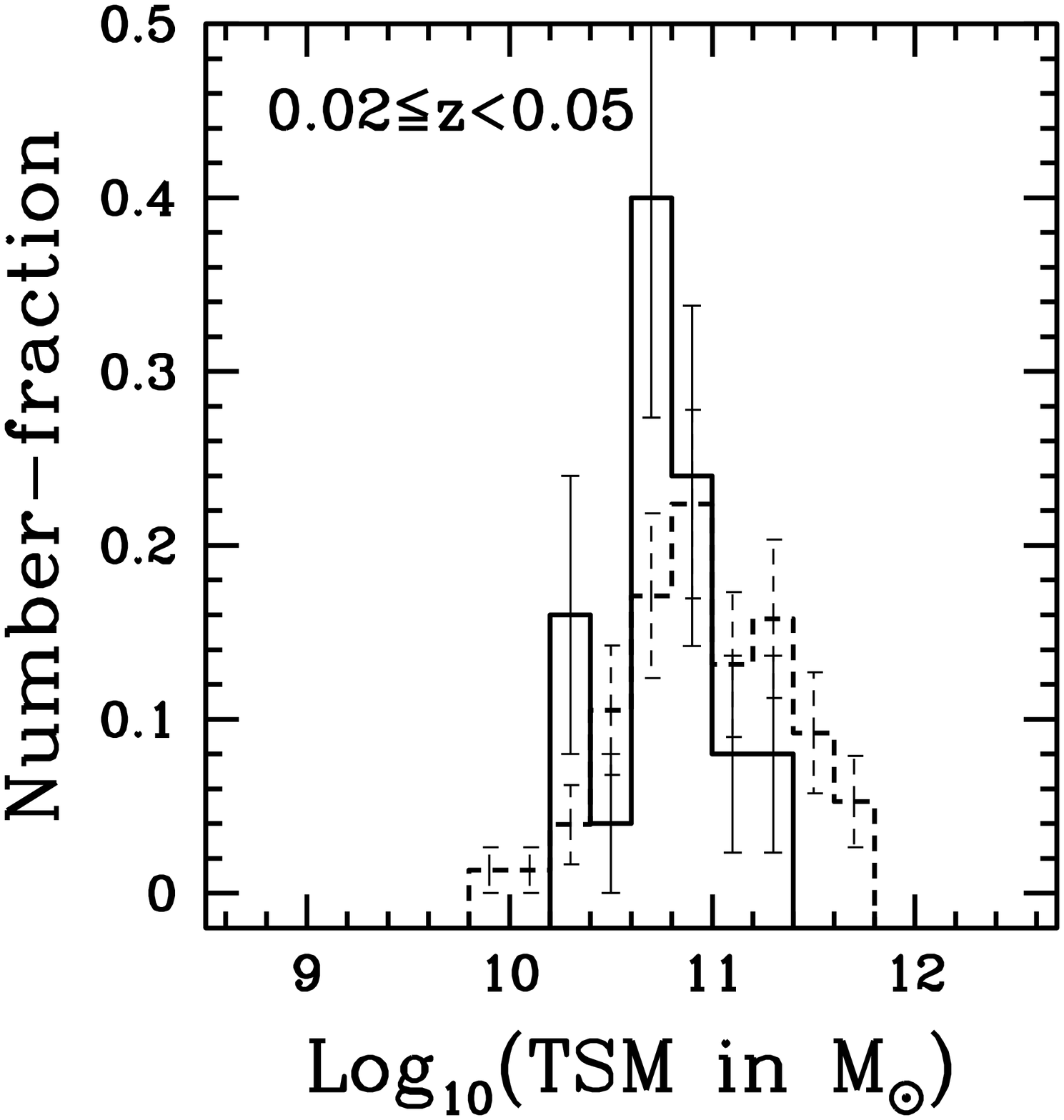}}
\mbox{\includegraphics[width=60mm]{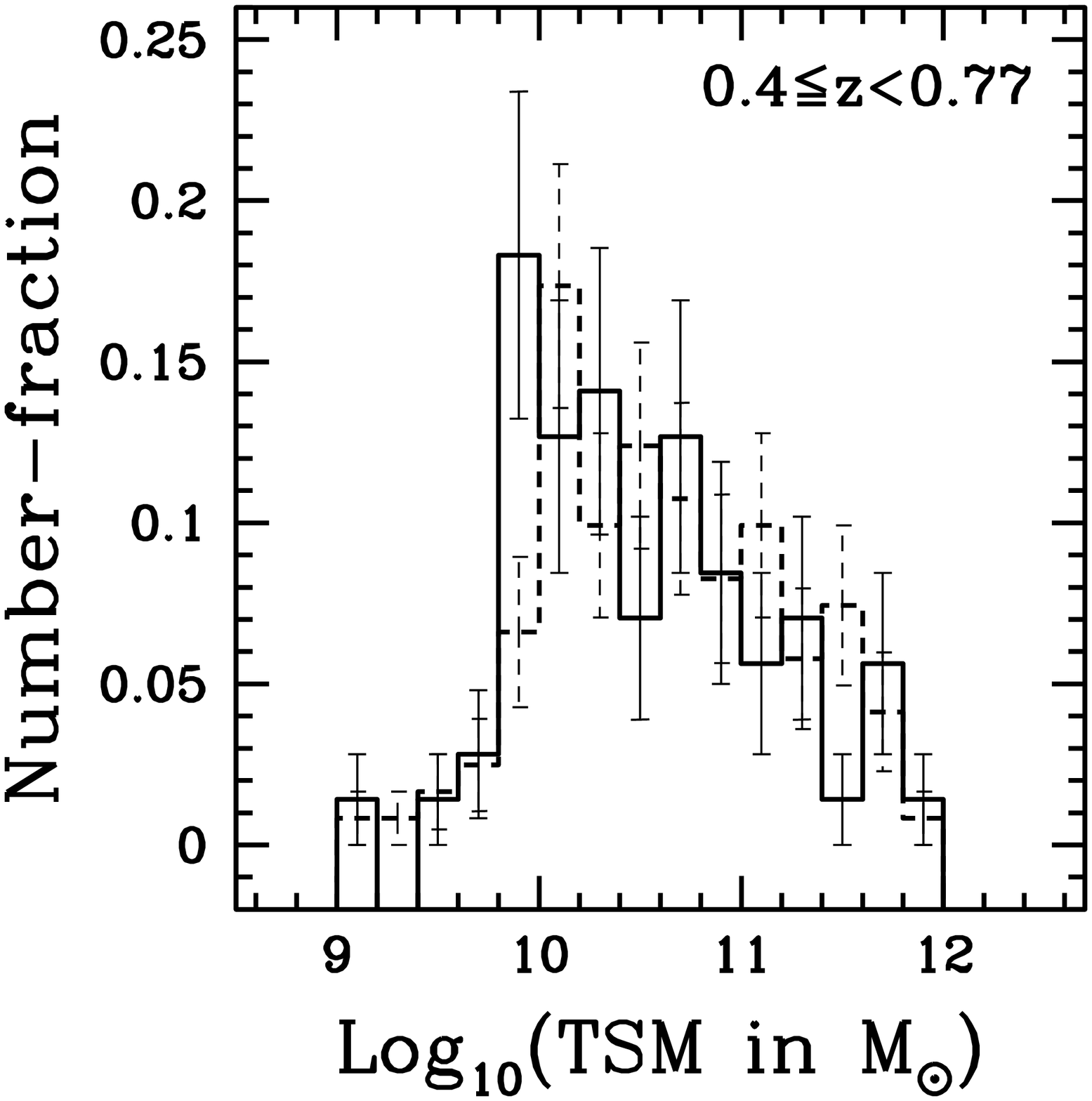}}
\mbox{\includegraphics[width=60mm]{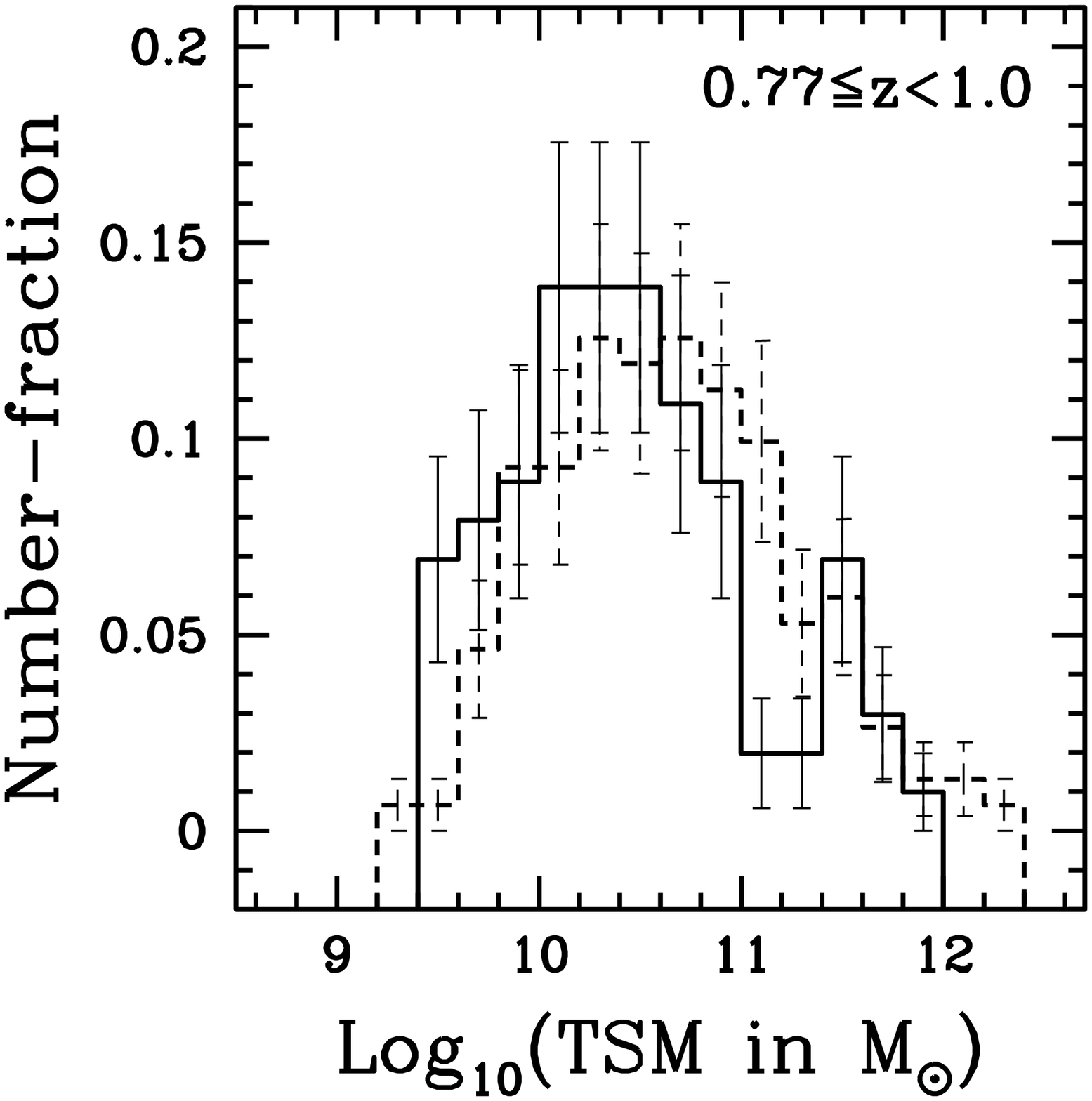}}
\mbox{\includegraphics[width=60mm]{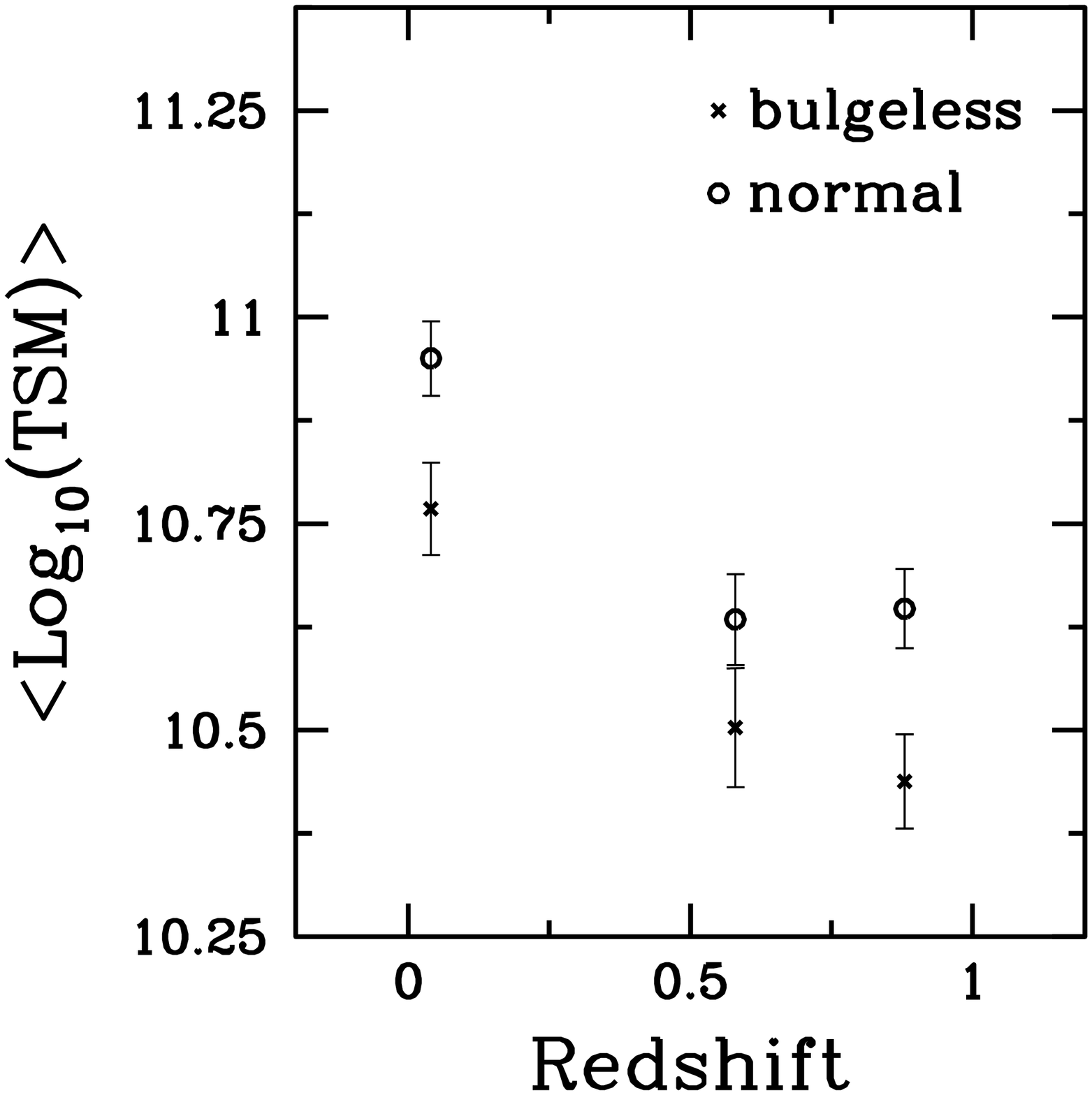}}
\caption{Distribution of log of total stellar mass (in units of M$_\odot$) is shown in the three redshift ranges for bulgeless disc (solid lines) and normal disc (dashed lines) galaxies. The distribution of their mean values with redshift is also shown. Since it is on log scale, even a slight shift corresponds to a huge increase in the mass of the galaxy. A shift towards higher mass can be seen for both morphological types. Their stellar mass at $z\sim0$ is more than double of that at $z\sim0.9$.}
\label{histltsm}
\end{figure*}

Total stellar mass is one of the most significant properties of a galaxy and is seen to be correlated with not just the overall concentration but also the star formation rate of the galaxy \citep{Caonetal1993,Conselice2003,Noeskeetal2007,Disneyetal2008,Baueretal2011}. Recent studies have found that close to half of the present stellar mass of the galaxies assembled by $z\sim1$ \citep{Bundyetal2005,Mortlocketal2011,Marchesinietal2014,Ownsworthetal2014}. We first examine the growth in the stellar mass for our full sample from $z\sim0.9$ to the present epoch (Fig.~\ref{histltsm}).

The mean values of the log of total stellar mass (in units of solar mass, M$_\odot$) are given in the three redshift ranges in Table~\ref{meanltsmlesmd}. The bulgeless disc galaxies witness an increase of 3.1$\times$10$^{10}$ M$_\odot$ (from 2.7 to 5.8$\times$10$^{10}$ M$_\odot$) from $z\sim0.9$ to the present epoch. Over the same time range, the normal disc galaxies witness an increase of 4.5$\times$10$^{10}$ M$_\odot$ (from 4.4 to 8.9$\times$10$^{10}$ M$_\odot$). 

Thus, we find that both the morphological types have gained more than 50($\pm$6)\% of their present stellar mass since $z\sim0.9$. The interesting part is that the increase in the total stellar mass of a normal disc galaxy is $\sim$1.5 times more than that seen for a bulgeless disc galaxy over the last $\sim$8 Gyrs, on average. The difference in the average total stellar mass of the two morphological types almost doubles (from 1.7 to 3.1$\times$10$^{10}$ M$_\odot$) from $z\sim0.9$ to the present epoch. 
 
\begin{figure*}
\mbox{\includegraphics[width=60mm]{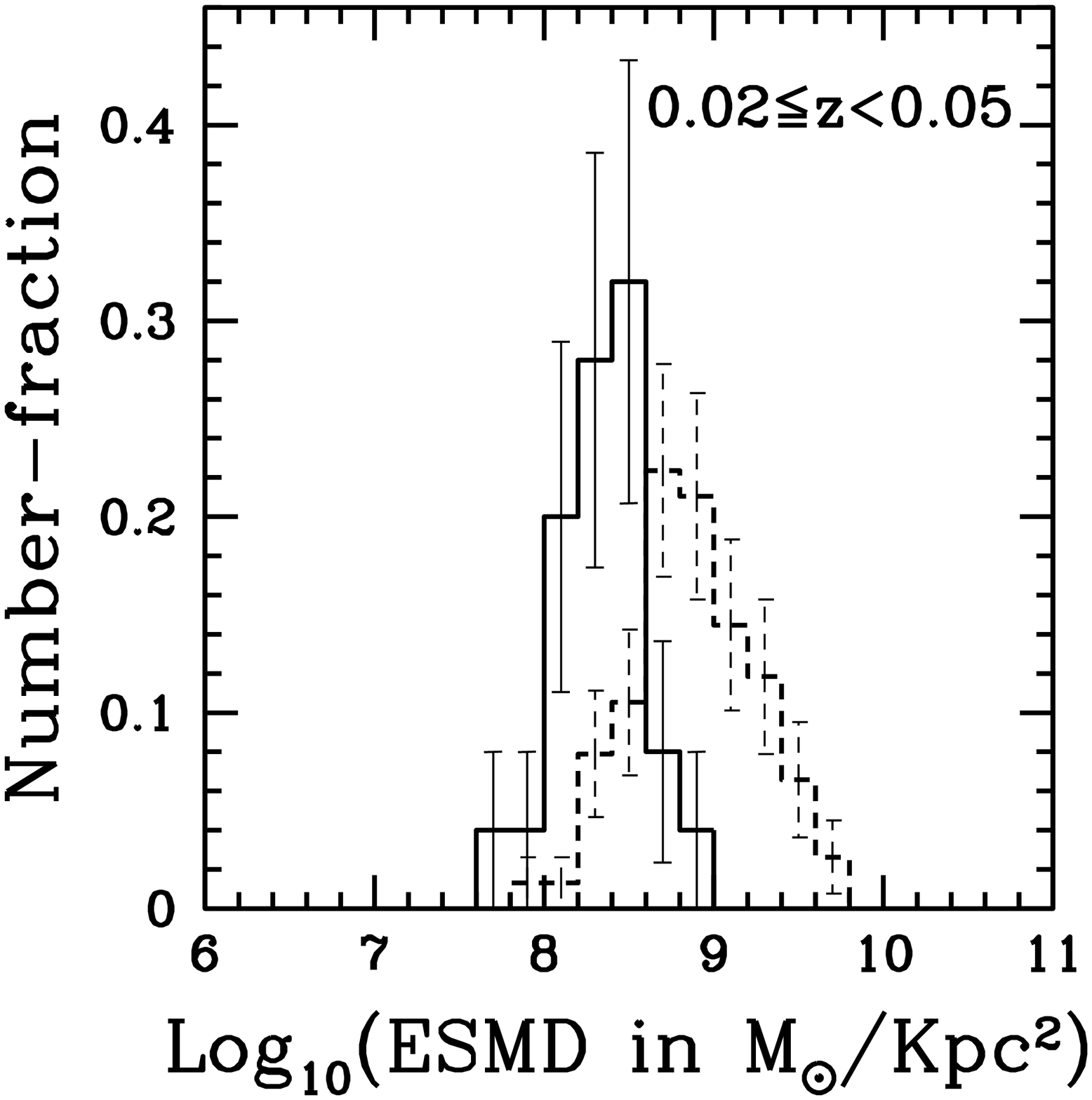}}
\mbox{\includegraphics[width=60mm]{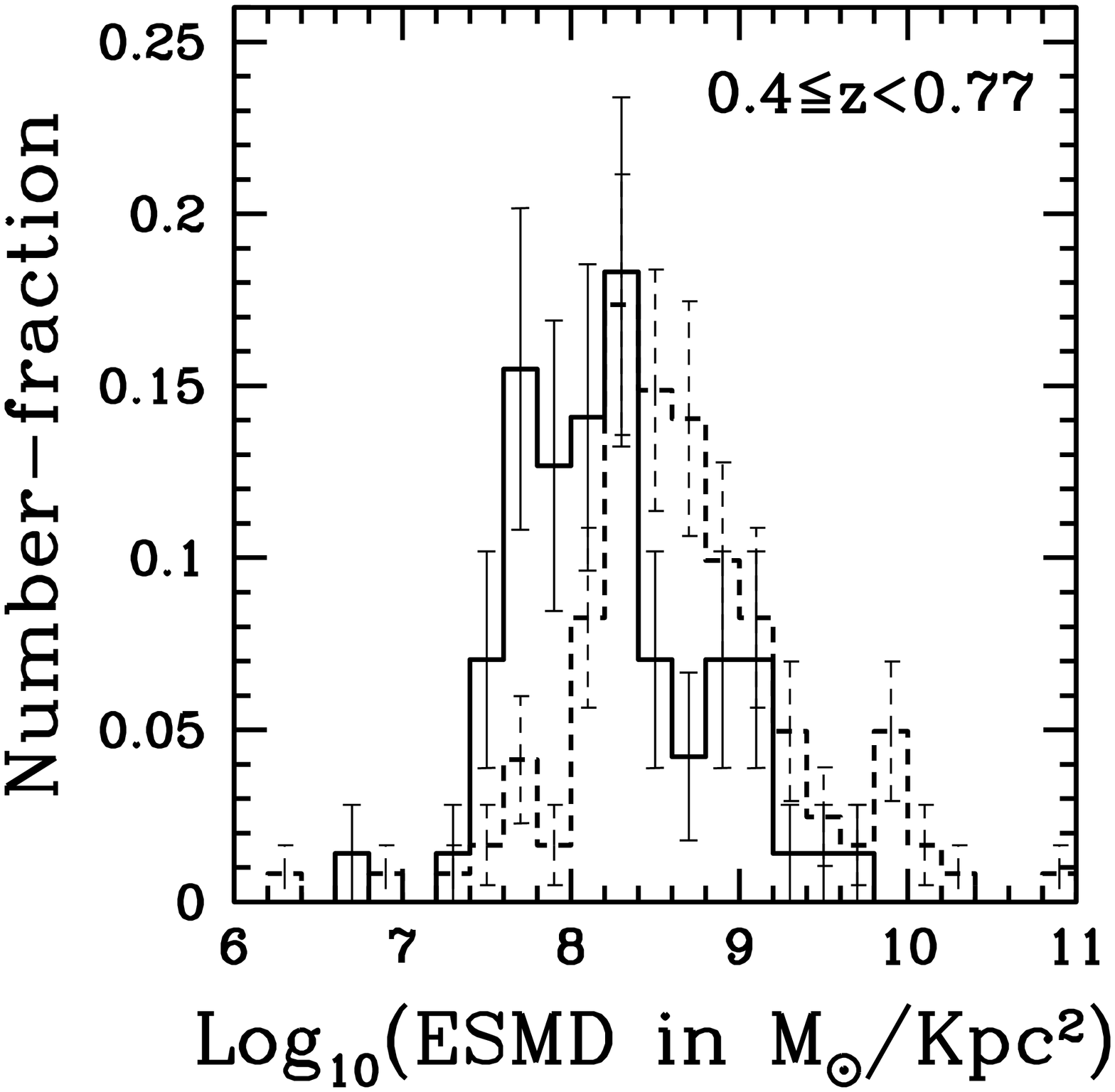}}
\mbox{\includegraphics[width=60mm]{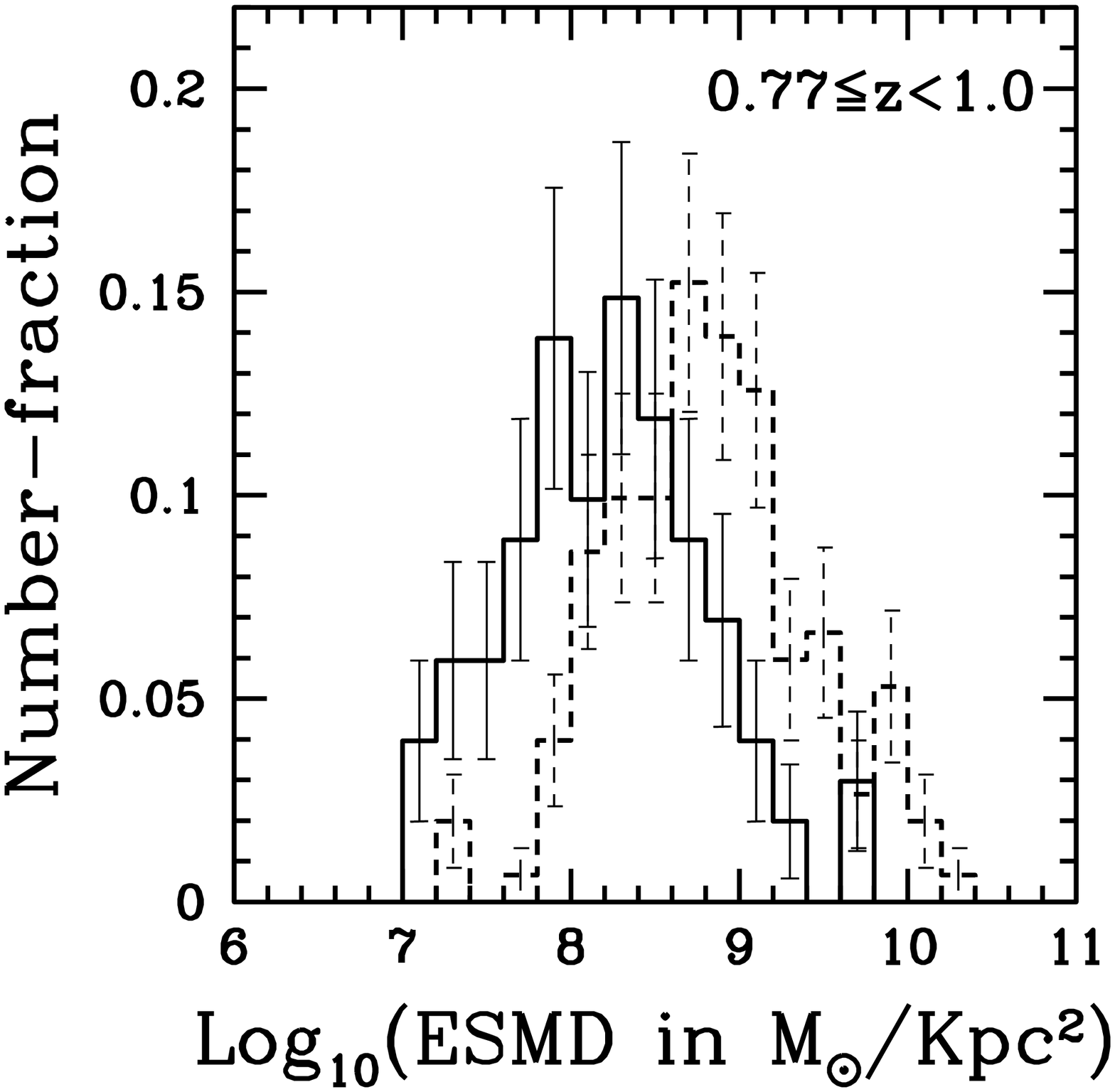}}
\mbox{\includegraphics[width=60mm]{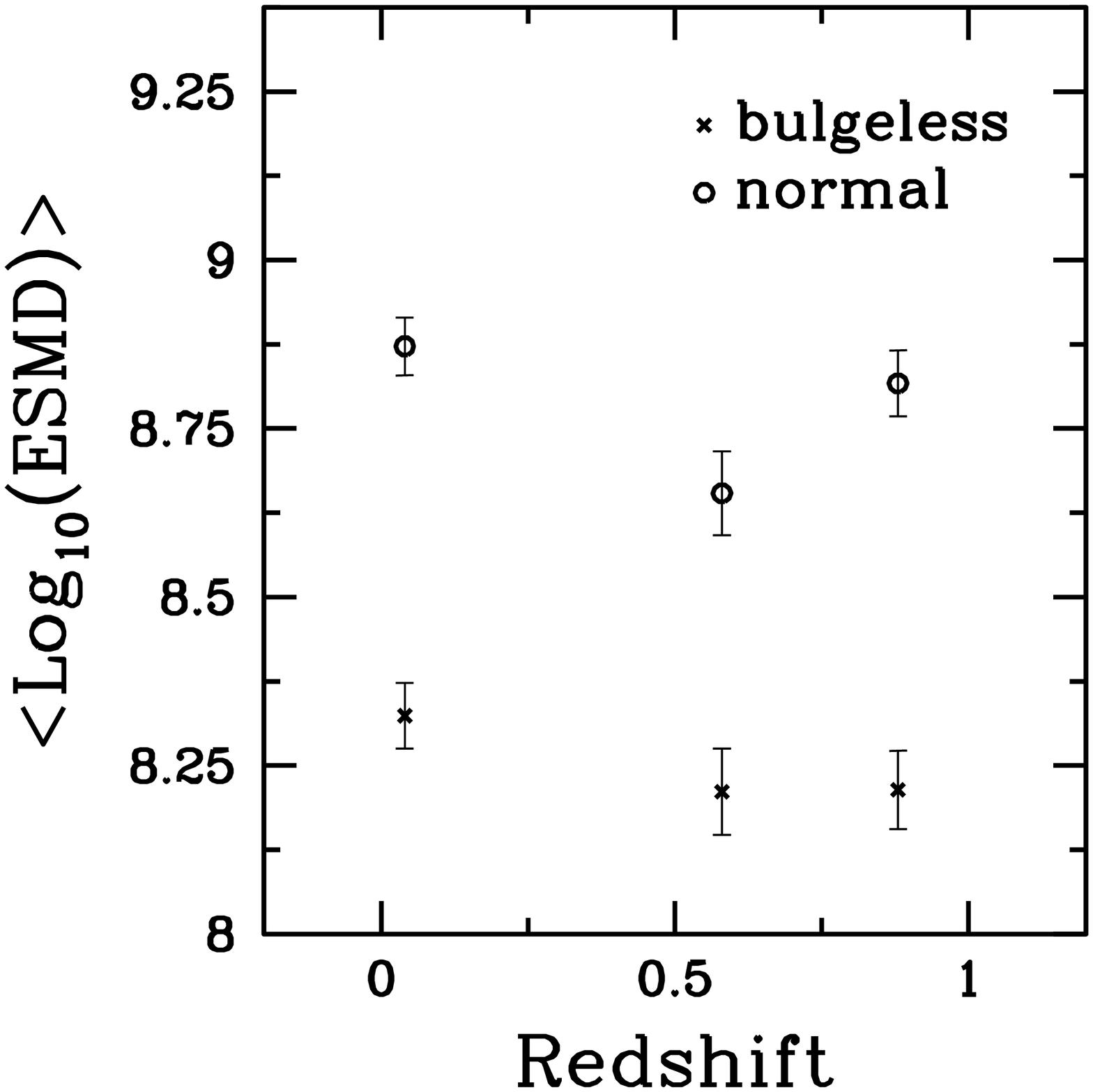}}
\caption{Distribution of log of effective stellar mass density is shown in the three redshift ranges for bulgeless disc (solid lines) and normal disc (dashed lines) galaxies. The distribution of their mean values with redshift is also shown. Since it is on log scale, even a slight shift corresponds to a huge increase in the mass density of the galaxy. The normal sample shows fluctuation, however, the density at $z\sim0$ is more than that at $z\sim0.9$ for both morphological types.}
\label{histlesmd}
\end{figure*}

Next we examine the growth of stellar mass density in the inner region i.e. the effective stellar mass density. The distribution of the effective stellar mass density on the log-scale is shown in Fig.~\ref{histlesmd} for bulgeless and normal disc galaxies in the three redshift ranges. 

Examining the mean values from $z\sim0.9$ to the present epoch (Table~\ref{meanltsmlesmd}), we find an increase of 4.7$\times$10$^{7}$ M$_\odot$/kpc$^2$ (from 1.6 to 2.1$\times$10$^{8}$ M$_\odot$/kpc$^2$), on average, for the effective stellar mass density of bulgeless disc galaxies; and of 8.8$\times$10$^{7}$ M$_\odot$/kpc$^2$ (from 6.6 to 7.4$\times$10$^{8}$ M$_\odot$/kpc$^2$), on average, for the normal disc galaxies. 

While the normal disc galaxies witness an increase of 13($\pm$5)\% in their effective stellar mass density, this is more prominent for the bulgeless disc galaxies, namely 30($\pm$1)\%. However, in absolute terms, the increase for the average normal disc galaxy is $\sim$1.8 times that seen for a bulgeless disc galaxy over the last $\sim$8 Gyrs. Thus, for both total stellar mass and effective stellar mass density, the increase seen for the normal disc sample is considerably more than that seen for the bulgeless disc sample.

In addition, we note that the effective stellar mass density can be a better indicator of galaxy morphology than concentration. This is seen using the Kolmogorov-Smirnov test which is considered an efficient mathematical tool for determining the significance of the difference of two distributions. We use this test to compare the bulgeless and normal disc galaxy samples with respect to concentration and effective-mass-density in three redshift ranges (Table~\ref{kstest}). 

The null hypothesis is that the two distributions are from the same parent set and this test quantifies a probability for this null hypothesis. For that, we compute the distance (D-observed) between the empirical cumulative distribution functions of the two samples. Based on this distance and the sample size, the probability for the null hypothesis is found (Table~\ref{kstest}). In the case of concentration, this probability is not convincingly low for the two distributions to be considered significantly different. However, it is found to be negligible at all redshift ranges in the case of effective stellar mass density.

\begin{table*}
\begin{minipage}{200mm}
\caption{Mean values of log of total stellar mass ($LTSM$) and log of effective-stellar-mass-density ($LESMD$).}
\begin{tabular}{@{}llllllllll@{}}
\hline
Redshift & Disc & No. of & Mean of & Std.dev. & Median & Mean of & Std.dev. & Median\\
range & type & sources & $LTSM$ & of $LTSM$ & of $LTSM$ & $LESMD$ & of $LESMD$ & of $LESMD$\\
 & & & $<LTSM>$ & $\sigma$ & & $<LESMD>$ & $\sigma$ &\\
\hline
0.77-1.0 & bulgeless & 101 & 10.438($\pm$0.057) & 0.574 & 10.381($\pm$0.071) & 8.214($\pm$0.058) & 0.584 & 8.219($\pm$0.073)\\
0.77-1.0 & normal & 151 & 10.647($\pm$0.048) & 0.593 & 10.610($\pm$0.060) & 8.817($\pm$0.049) & 0.598 & 8.795($\pm$0.061)\\
0.4-0.77 & bulgeless & 71 & 10.503($\pm$0.072) & 0.603 & 10.386($\pm$0.090) & 8.211($\pm$0.064) & 0.544 & 8.178($\pm$0.080)\\
0.4-0.77 & normal & 121 & 10.634($\pm$0.055) & 0.609 & 10.562($\pm$0.069) & 8.654($\pm$0.062) & 0.682 & 8.575($\pm$0.078)\\
0.02-0.05 & bulgeless & 25 & 10.768($\pm$0.056) & 0.280 & 10.755($\pm$0.070) & 8.324($\pm$0.049) & 0.248 & 8.339($\pm$0.061)\\
0.02-0.05 & normal & 76 & 10.950($\pm$0.045) & 0.391 & 10.943($\pm$0.056) & 8.872($\pm$0.043) & 0.377 & 8.854($\pm$0.054)\\
\hline
\label{meanltsmlesmd}
\end{tabular}
\end{minipage}
\end{table*}

\begin{table}
\centering
\begin{minipage}{100mm}
\caption{KS-test D-observed and Probability values}
\begin{tabular}{@{}llllllllll@{}}
\hline
Parameter & Redshift & D-observed & Probability\\
 & range & &\\
\hline
Conc & 0.02-0.05 & 0.337 & 0.021\\
Log-EMD & 0.02-0.05 & 0.709 & 3.4e-09\\
Conc & 0.40-0.77 & 0.178 & 0.092\\
Log-EMD & 0.40-0.77 & 0.395 & 9.7e-07\\
Conc & 0.77-1.00 & 0.132 & 0.204\\
Log-EMD & 0.77-1.00 & 0.414 & 8.9e-10\\
\hline
\label{kstest}
\end{tabular}
\end{minipage}
\end{table}

\subsection{Correlations with asymmetry}
The asymmetric features in the disc galaxy, created by both internally and externally driven processes, lead to the formation of pseudo- and classical bulges in disc galaxies \citep{KhochfarandSilk2006,Elmegreenetal2008,Jogeeetal2009,Kormendyetal2010,Hopkinsetal2010,Conselice2014}. Thus, it is imperative to analyse the evolution of asymmetry and its relationship with other parameters of disc galaxies.

\begin{figure*}
\mbox{\includegraphics[width=60mm]{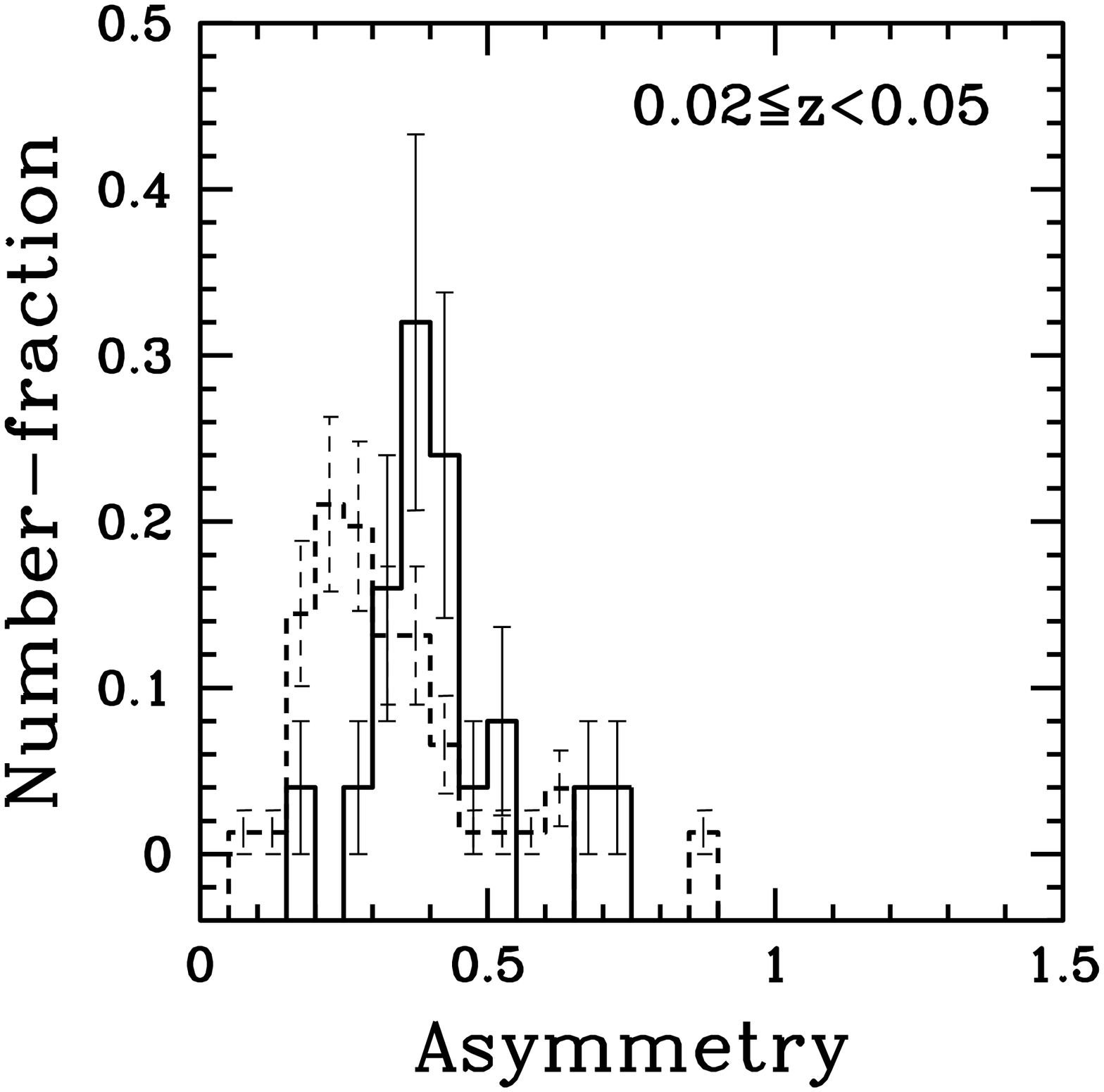}}
\mbox{\includegraphics[width=60mm]{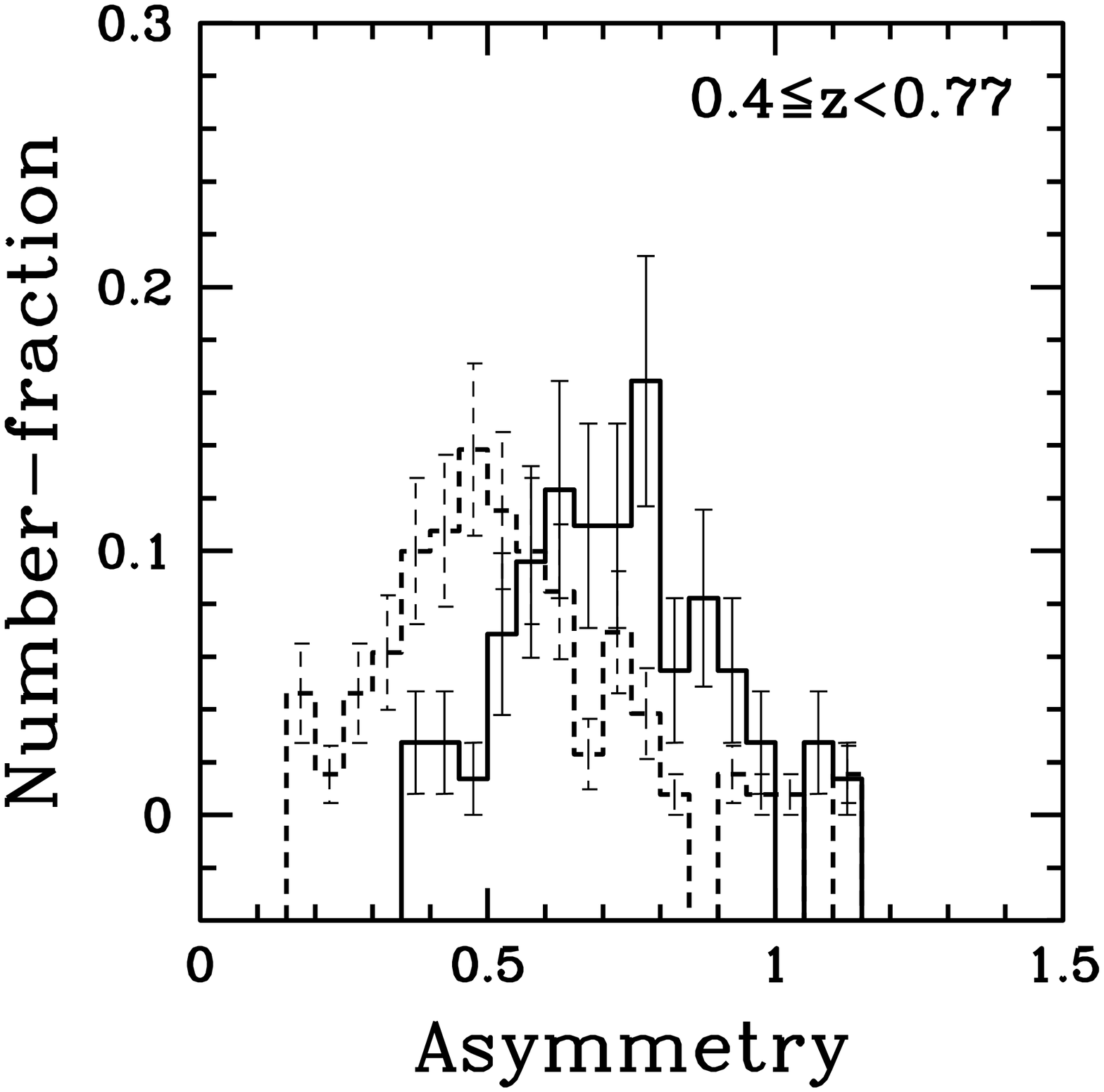}}
\mbox{\includegraphics[width=60mm]{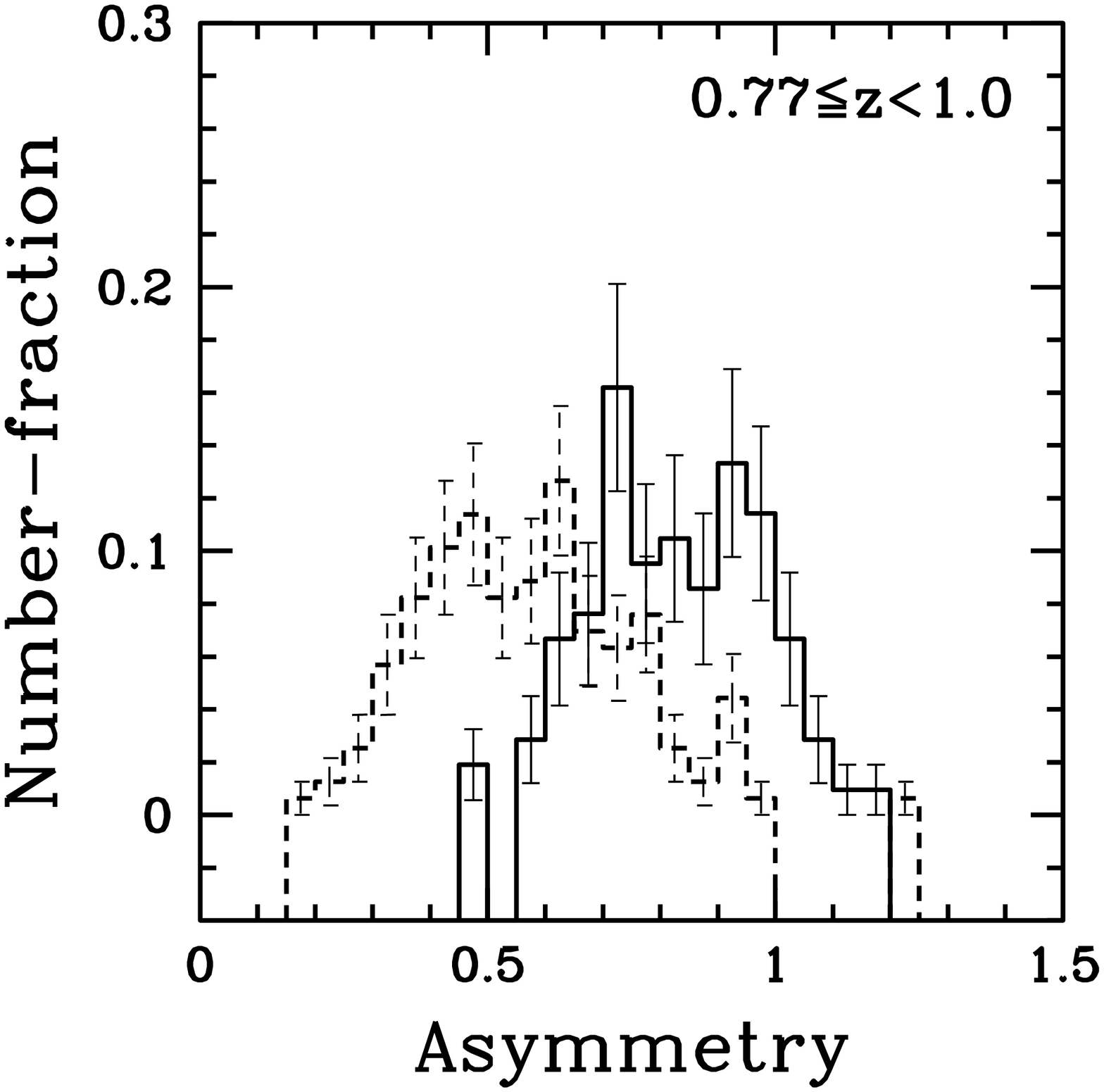}}
\mbox{\includegraphics[width=60mm]{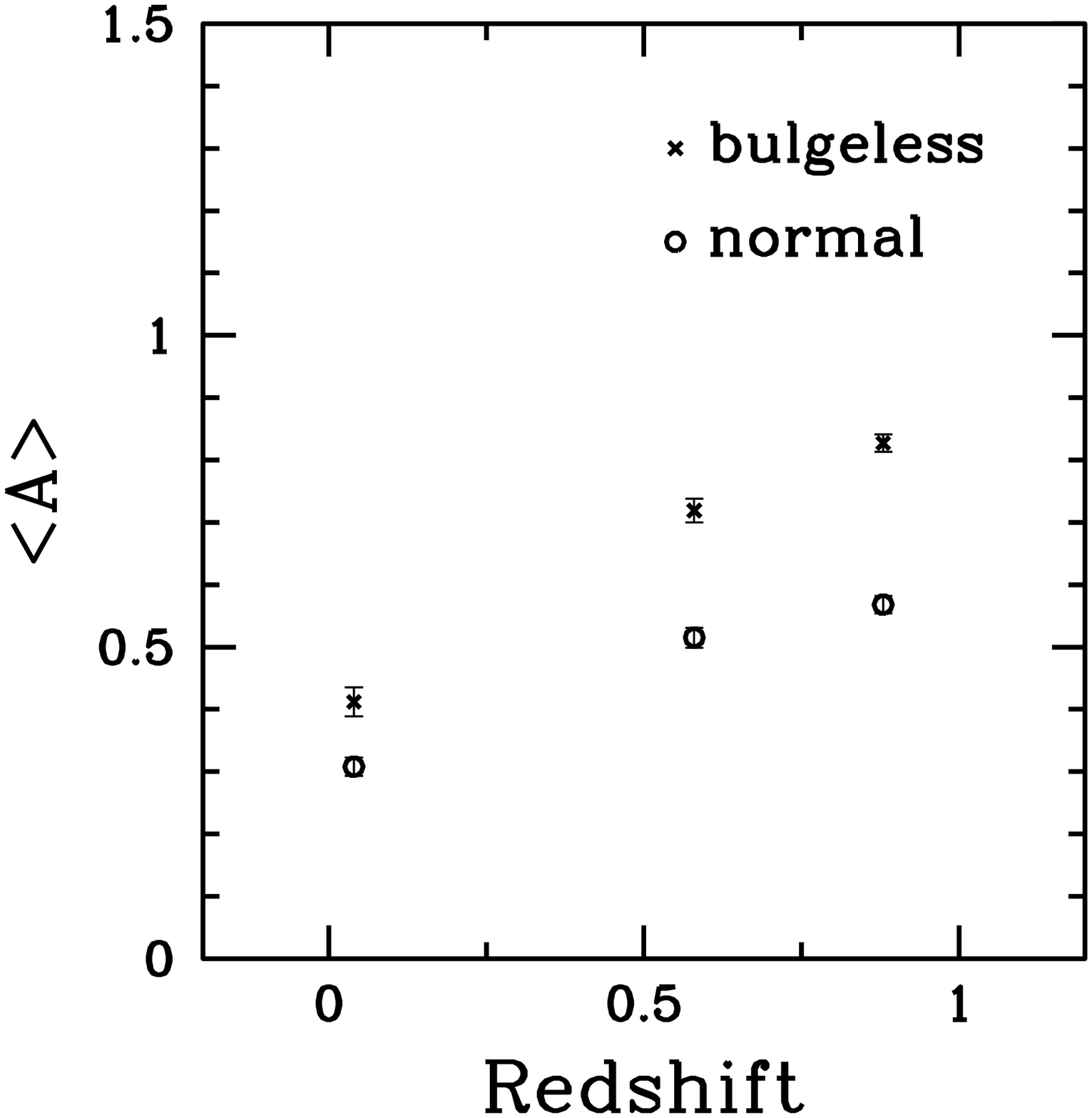}}
\caption{Distribution of asymmetry for bulgeless disc (solid lines) and normal disc (dashed lines) galaxies is shown for the three redshift ranges. The distribution of the means with redshift is also shown. Reduction in the scatter of the parameter value with time is apparent. The fall in the mean value of asymmetry is more significant for the bulgeless disc sample.}
\label{histasym}
\end{figure*}

The distribution of the asymmetry for bulgeless and normal disc galaxies is shown in Fig.~\ref{histasym} for the three redshift ranges. The mean asymmetry for bulgeless disc galaxies falls by 50($\pm$3)\% (from 0.83 to 0.41) from $z\sim0.9$ to the present epoch (Table~\ref{meanconcasym}). Over the same time range, the mean asymmetry for the normal disc sample falls by 45($\pm$2)\% (from 0.57 to 0.31; Table~\ref{meanconcasym}).

Both bulgeless and normal disc galaxies show a huge decline in their asymmetry value with time, indicating the disappearance of asymmetric features. The bulgeless disc sample is more asymmetric than the normal disc sample at all redshift ranges (Fig.~\ref{histasym}). However, due to the more significant fall seen in the average asymmetry value of the bulgeless disc sample, it reaches closer to the average asymmetry value of the normal disc sample.

\begin{figure*}
\mbox{\includegraphics[width=55mm]{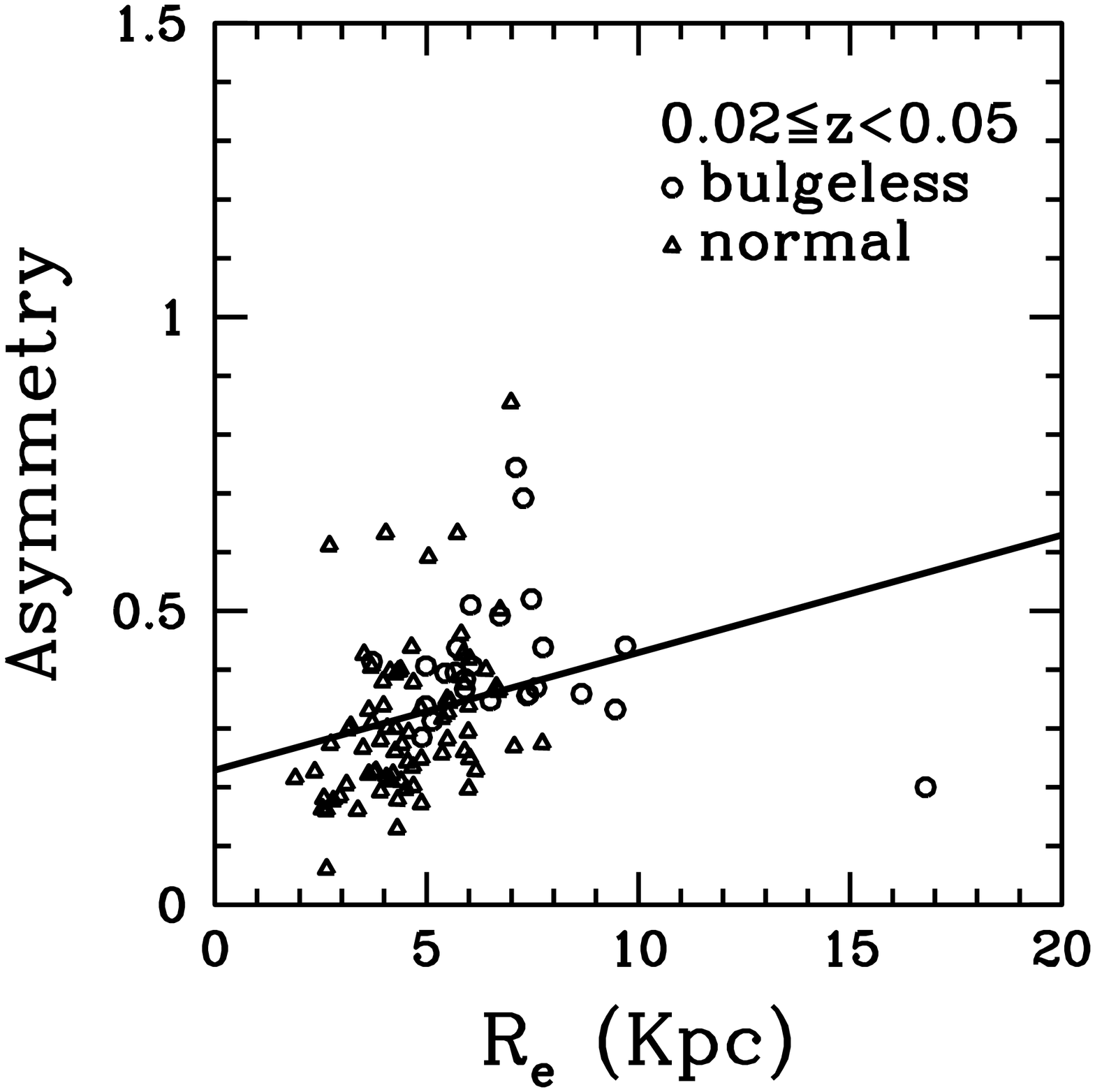}}
\mbox{\includegraphics[width=55mm]{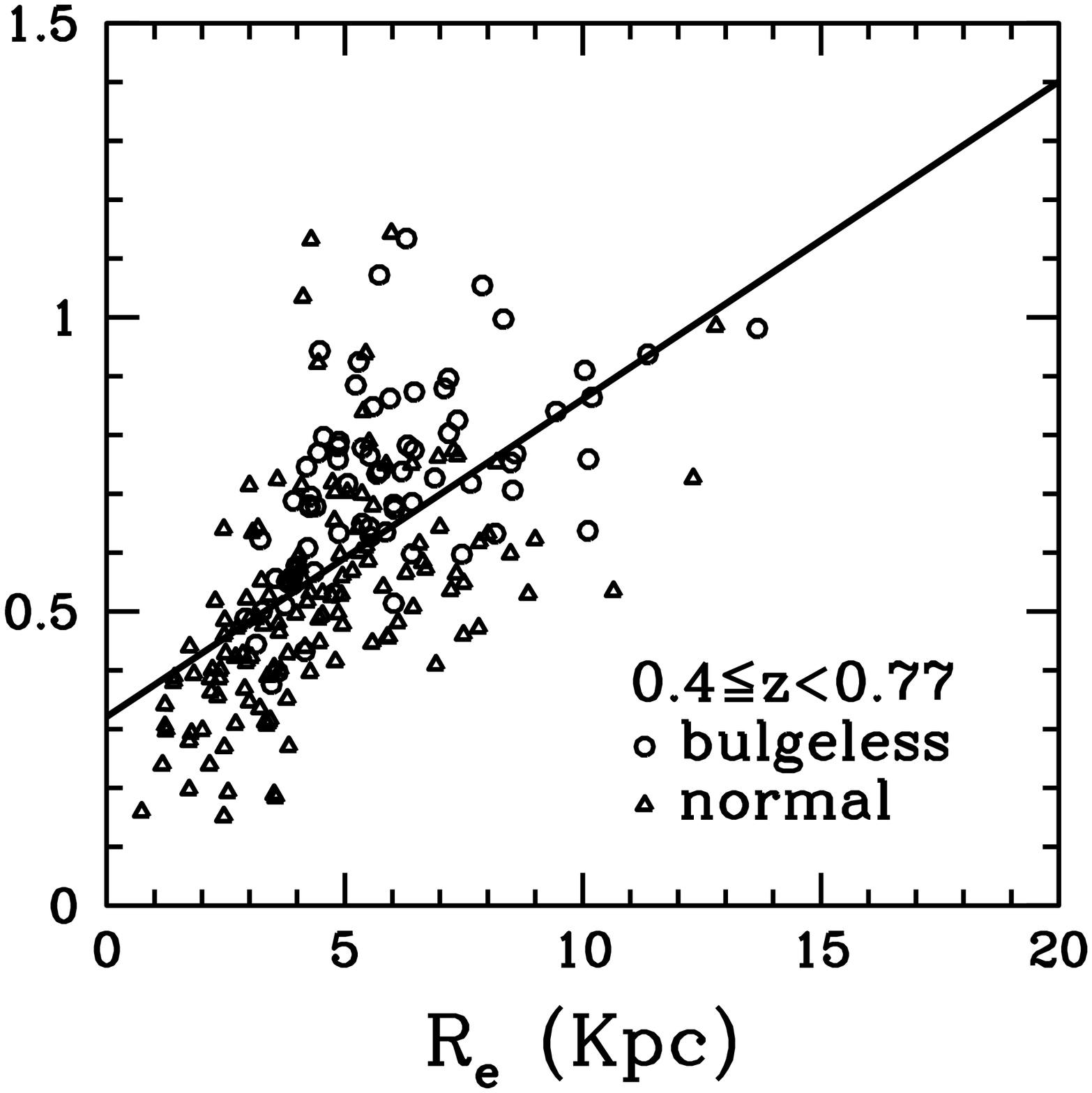}}
\mbox{\includegraphics[width=55mm]{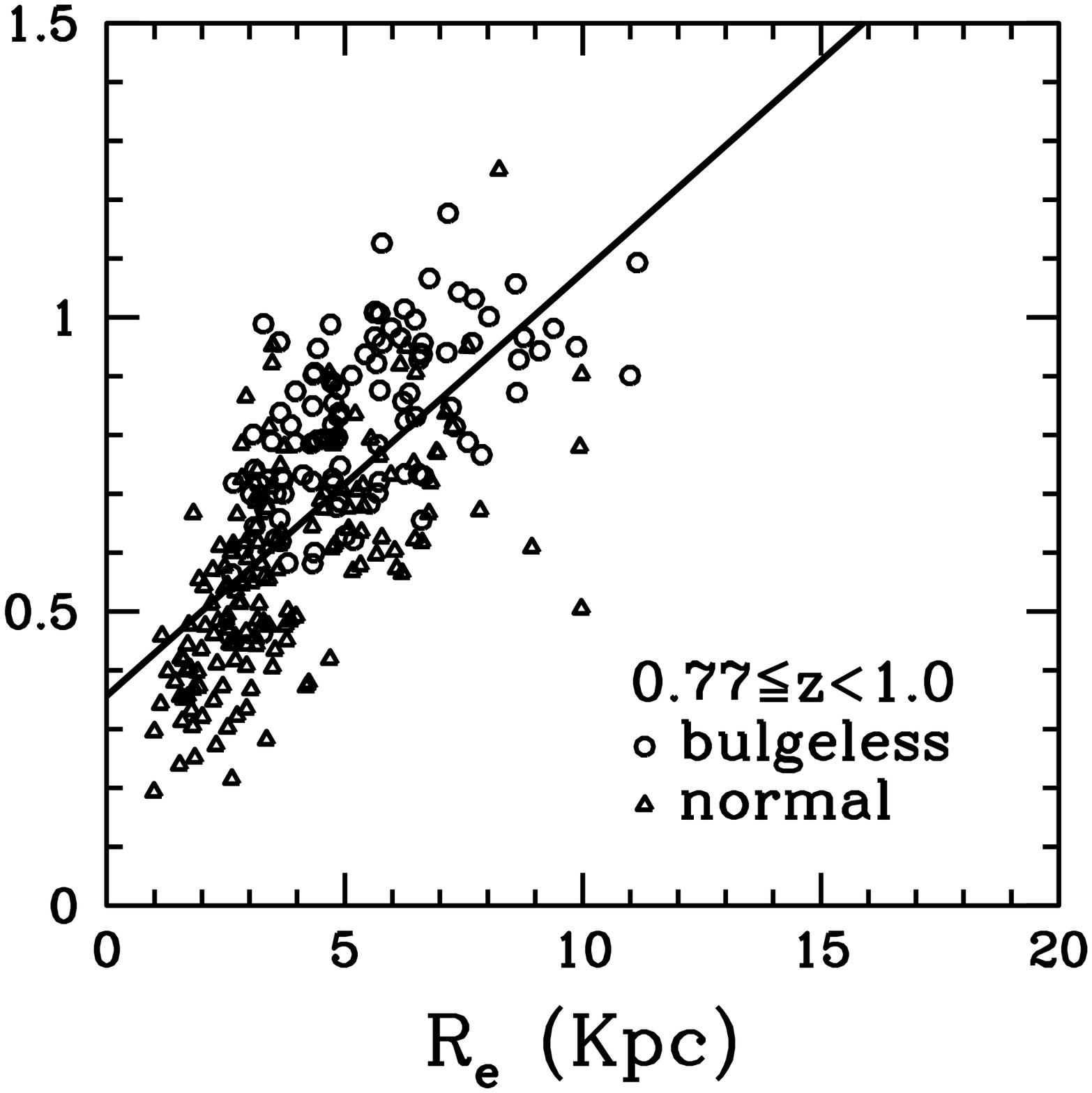}}
\caption{Asymmetry is plotted against half light radius for bulgeless disc and normal disc galaxies in the three redshift ranges. The solid line in each graph marks the linear relation followed by the full sample at that redshift range. Larger the radius, greater is seen to be the asymmetry value of the galaxy. Sources from the bulgeless disc sample, having on an average larger half light radii, are found to have higher asymmetry values. The typical error on the asymmetry value is $\pm$0.08 and for the half light radius it is $\pm$0.2 kpc.}
\label{asymvsre}
\end{figure*}

The asymmetry parameter shows a correlation with half-light radius (Fig.~\ref{asymvsre}). Bulgeless disc galaxies, having larger half-light radii than the normal disc galaxies, have higher asymmetries. At fixed radii, bulgeless disc galaxies are found to be more asymmetric than normal disc galaxies, on average.

\begin{figure*}
\mbox{\includegraphics[width=55mm]{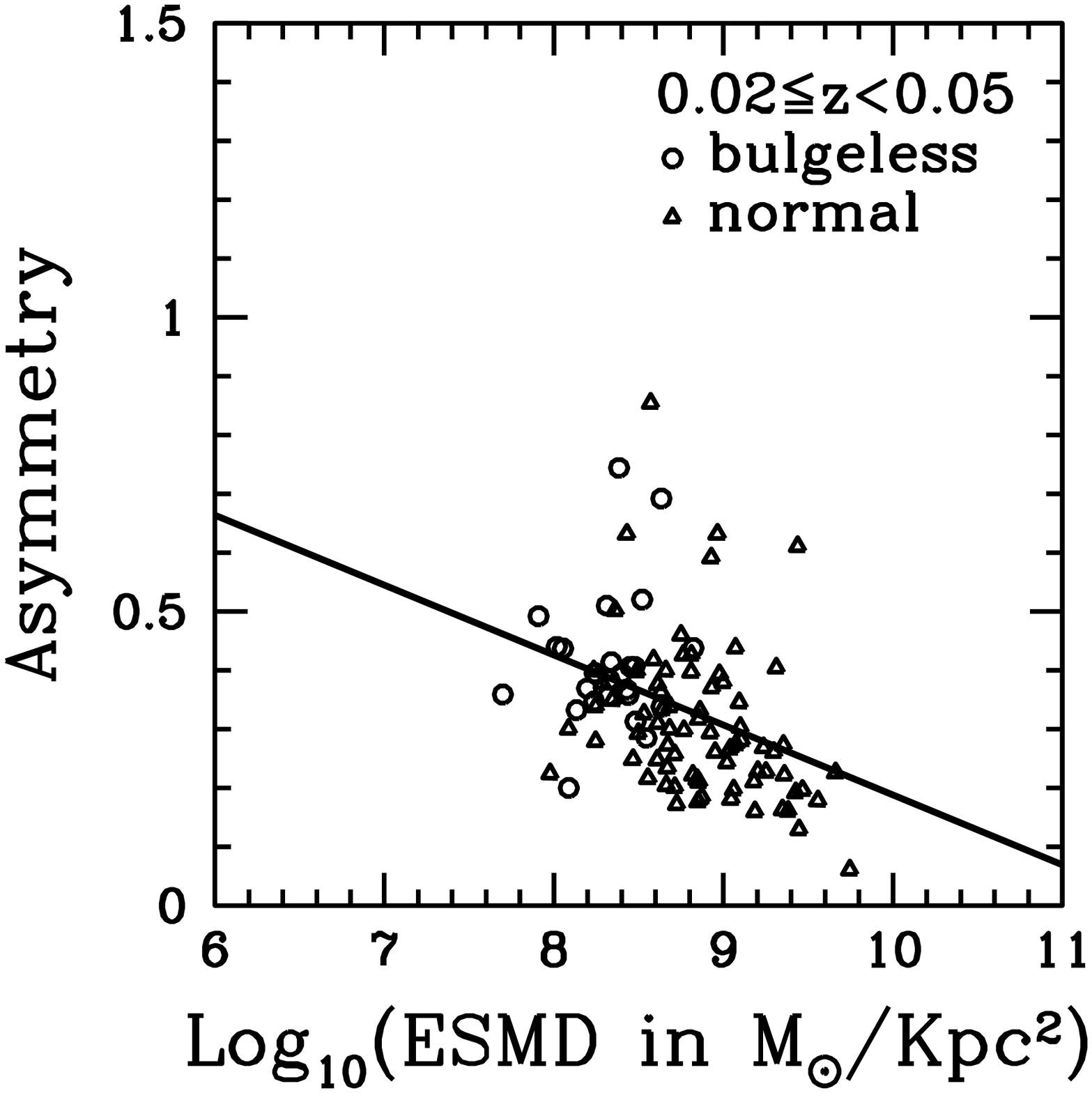}}
\mbox{\includegraphics[width=55mm]{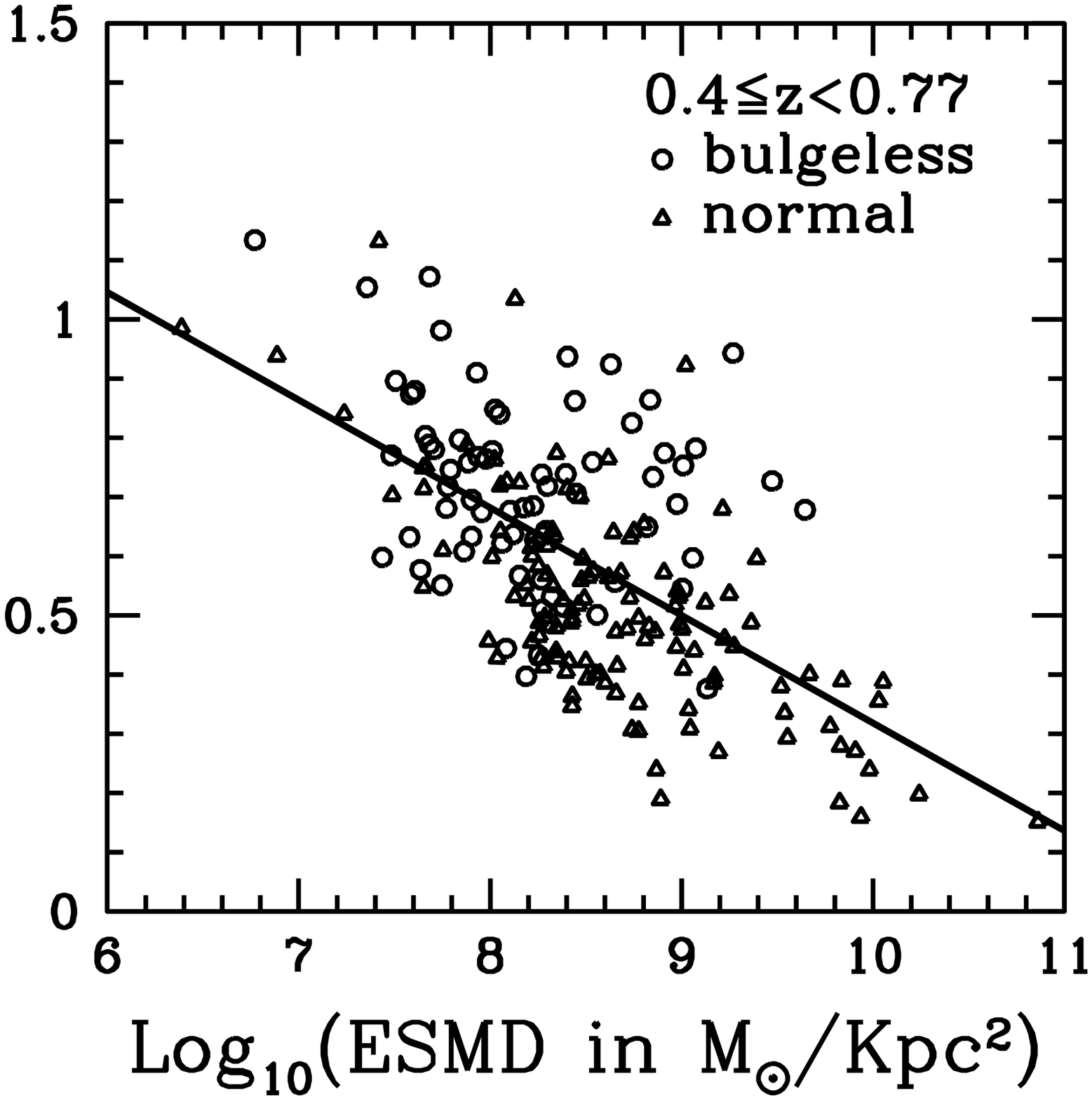}}
\mbox{\includegraphics[width=55mm]{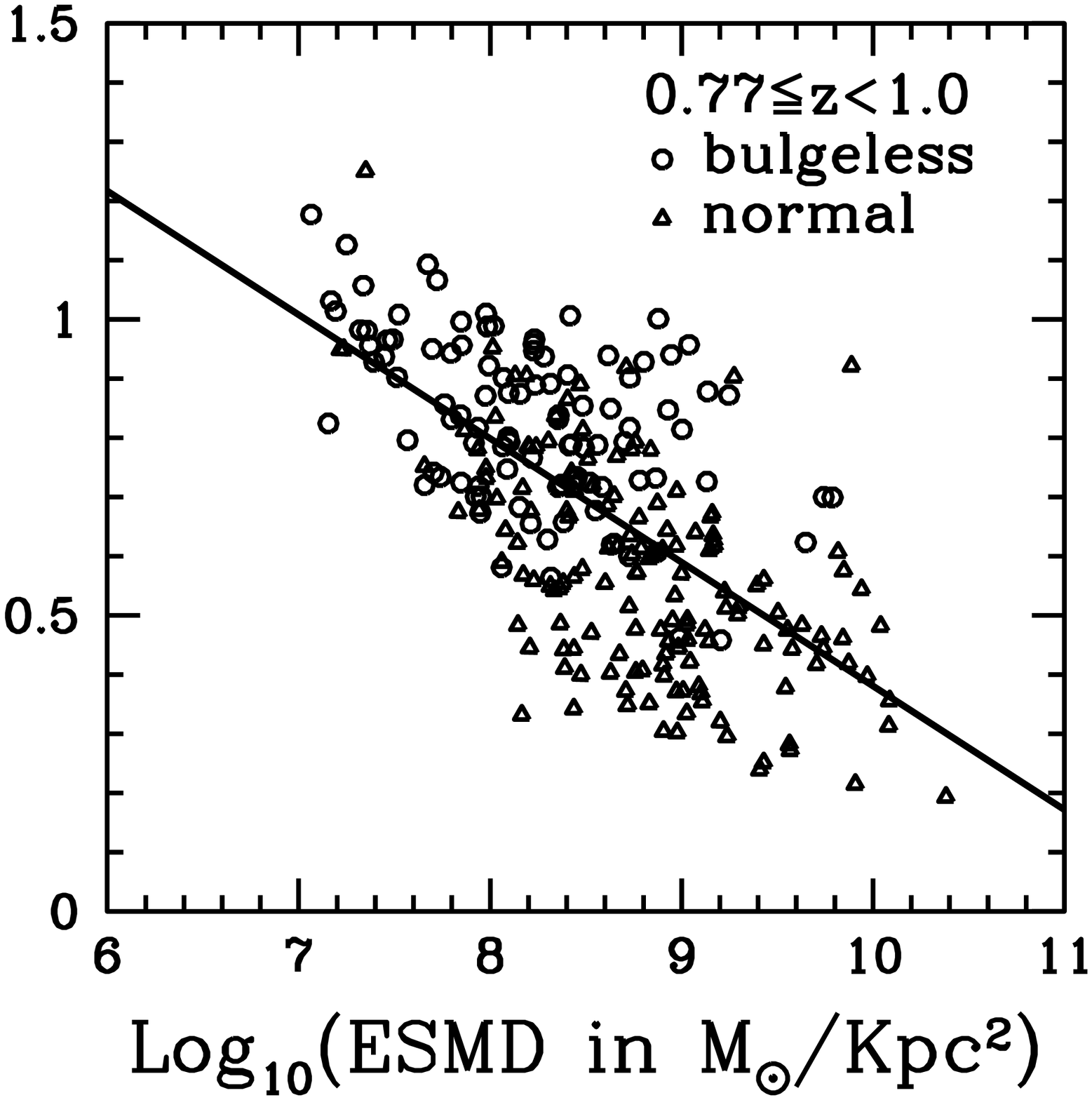}}
\caption{Asymmetry is plotted against log of effective stellar mass density for bulgeless disc and normal disc galaxies in the three redshift ranges. The solid line in each graph marks the linear relation followed by the full sample at that redshift range. The two parameters are seen to be highly correlated for both morphological types. Higher the galaxy's stellar mass density inside its effective radius, lower is the asymmetry of its stellar light distribution. The typical error on the asymmetry value is $\pm$0.08 and for the log of effective stellar mass density it is $\pm$0.05.}
\label{asymvslesmd}
\end{figure*}

The asymmetry parameter shows an anti-correlation with effective stellar mass density, as explored in Fig.~\ref{asymvslesmd} for both morphological types. There is a steep fall in the asymmetry with the increase in effective stellar mass density for all the redshift ranges. The slope for 0.4$\leq z<$1.0 is:
\begin{equation}
A=-0.21(\pm0.02)*log(EMD)+2.47(\pm0.13).
\end{equation}

Bulgeless disc galaxies being more asymmetric and less dense as compared to normal disc galaxies are on the higher end of the slope in all three redshift ranges. 

\section{Discussion}
We have a total sample of 567 disc-dominated galaxies separated into bulgeless (without classical bulge) discs and normal (with classical bulge) discs in three redshift ranges (0.77$\leq z<$1.0, 0.4$\leq z<$0.77 and 0.02$\leq z<$0.05). We have examined the evolution in size, concentration, effective stellar mass density and asymmetry of these two samples. We first list our major findings and then discuss the implications:
 
\begin{enumerate}
\item Both morphological types show a significant increase [94($\pm$6)\% for the normal disc sample] in total optical extent since $z\sim0.9$. The half-light radius of the galaxies witnesses a much smaller increase [22($\pm$1)\% for the normal disc sample] in comparison. This peripheral size evolution is more evident for the normal disc sample in which the outer radius becomes $\sim$5 times the inner radius by $z\sim0$.
\item The mean concentration of stellar light undergoes a significant increase ($\Delta$$C=0.34$) for the normal disc sample as compared to the bulgeless disc sample ($\Delta$$C=0.04$) since $z\sim0.9$. The concentration parameter of the galaxy is seen to be well correlated with its global S\'ersic index value for both morphological types at higher redshift ranges ($0.44-0.77$, $0.77-1.0$). However, the correlation is absent for the local sample ($0.02-0.05$).
\item Both morphological types have gained more than half of their present stellar mass since $z\sim0.9$. In absolute terms, the increase in the total stellar mass as well as effective stellar mass density is significantly more important ($\sim$1.5 and 1.8 times respectively) for the normal disc sample.
\item The bulgeless disc sample is more asymmetric than the normal disc sample at all redshift ranges. Both samples witness a fall in their mean asymmetry value from $z\sim0.9$ to $z\sim0$, the fall being more drastic [$\sim$50($\pm$3)\%] for the bulgeless disc sample. Asymmetry is found to be correlated with the half-light radius, and anti-correlated with the effective stellar mass density of the galaxy.
\end{enumerate}

\subsection{Impact of internal evolution}
In our sample, from $z\sim0.9$ to $z\sim0$, the Petrosian radius increases more significantly than the effective radius. In other words, the strong increase in the total radius of the full sample is not reflected in its half-light radius. This peripheral increase appears to provide evidence in support of internal secular evolution in which, due to the outward transfer of angular momentum, galaxy disks are expected to expand on the outside and get more concentrated on the inside \citep{Tremaine1989,Combes2001,KormendyandKennicutt2004}. However, there are complexities with this explanation.

The asymmetric features such as spiral arms and bars, that induce and speed up internal secular evolution \citep{Jogee2006,KormendyandKennicutt2004,CoelhoandGadotti2011,Shethetal2012,Cheungetal2013} are known to be present in the bulgeless disc galaxies with greater propensity \citep{Buta2013}. In our study also the bulgeless disc sample is found to be more asymmetric than the normal disc sample at all three redshift ranges. This in turn might favour the build-up of central concentration and eventual fading of the asymmetric features with time \citep{KormendyandKennicutt2004,Athanassoula2005} in bulgeless galaxies. This can be the probable cause for a relatively steeper decrease in the asymmetry value of our bulgeless disc sample. 

Naively, thus, internal secular evolution is expected to be more efficient for the bulgeless disc sample. However, we find that the evolution in terms of size and increase in the density of the inner-region, as seen through concentration and effective stellar mass density, in absolute terms, is considerably more for the normal disc sample. This gives us an indication that internal secular processes are not the only evolution determining forces.

In addition to that, it is known through simulations and observations that the huge increase in disc galaxies' total optical extent is occurring due to stellar mass build up in the outer regions of these galaxies \citep{Cappellarietal2009,Newmanetal2010}. We also report growth in stellar masses by a factor of two since z$\sim$1 for both the morphological samples.

This increase in size in terms of stellar mass build-up as well as the increase in internal density cannot be achieved only through the rearrangement of mass and angular-momentum. Thus, in addition to the internal secular evolution, there might be other processes causing the disc evolution from z$\sim$0.9 to the present epoch. We argue that such inside-out growth might be driven by the transfer of matter from the circumgalactic environment to the galaxy, in the next section.

\subsection{Impact of external evolution}
Our findings that discs have grown by such large factors from $z\sim1$ to now argues that stellar discs are robust structures, difficult to be destroyed, and that catastrophic mergers are rather rare at the second half of the age of the universe. This is in agreement with recent studies, based on simulations and observations, which report a significant decline in the major merger rate with time \citep{Jogeeetal2009,Conseliceetal2009,Blucketal2012}.

However, there is continuous accretion of matter from the intergalactic medium, and minor mergers are also frequent \citep{Parryetal2009,Kaviraj2010,Lopez-Sanjuanetal2011,Blucketal2012}. We do observe that both bulgeless and normal disc galaxies have gained more than 50\% of their total stellar mass in the past $\sim$8 billion years. Although star formation within the galaxy can also lead to the increase in its stellar mass, its contribution is measured to be much less compared to that from minor mergers and accretion \citep{Ownsworthetal2012,Ownsworthetal2014,MadauandDickinson2014}.

During the last $\sim$8 Gyrs, galaxies are predicted to suffer a period of intense bombardment by minor satellites \citep{KhochfarandSilk2006,Hopkinsetal2009,Feldmannetal2010,Oseretal2010,QuilisandTrujillo2012}. These bombardments are expected to bring morphological changes in the disc population by building classical bulges and also giving rise to spheroidal galaxies \citep{Oeschetal2010,Hopkinsetal2010,vanderWeletal2011,Cameronetal2011,Weinzirletal2011,Lawetal2012,Buitragoetal2013}. For the full sample studied, we report a considerable increase in the density of the inner region through the measures of S\'ersic index, concentration as well as the effective stellar mass density for the past $\sim$8 Gyrs. 

This increase is also observed in the fact that the fraction of disc galaxies with classical bulges is increasing from $z\sim0.9$ to the present epoch. It increases from 60\% at the highest redshift range (0.77-1.0) to 64\% at the intermediate redshift range (0.4-0.77) and finally to 75\% at the local redshift range (0.02-0.05), indicating that some of the bulgeless discs are growing a classical bulge with time. Although internal secular mechanisms driven by disc instabilities also lead to the increase in inner density, these methods can only lead to the formation of pseudo-bulges \citep{BinneyandTremaine1987,KormendyandKennicutt2004,Kormendyetal2010,SahaandGerhard2013}. The evidence, thus, suggests that minor mergers and accretion are playing a significant role from $z\sim0.9$ to the present epoch. 

There is substantial literature that establishes the dependence of galaxy properties on local environmental density such that the higher the density of the local environment, the more massive, dense, early-type and non-star-forming is the galaxy \citep{Dressler1984,Gomezetal2003,BlantonandMoustakas2009,Scovilleetal2013}. In the specific case of discs, discs with classical bulges are rarely found in low density environments \citep{Kormendyetal2010}. Also, discs without classical bulges are expected to keep this way by being in more isolated environments \citep{PeeblesandNusser2010}. By that argument, at $z\sim1$, normal disc galaxies supporting a classical bulge and being denser, more massive and less star-forming are expected to be placed in denser environments as compared to the bulgeless disc galaxies.

There is observational evidence that galaxies in denser environments show a more rapid increase of galaxy size with redshift \citep{Cooperetal2012,Lanietal2013}. Our results are in agreement, such that, normal disc galaxies show faster size evolution as compared to the bulgeless galaxies. Dense environments should also facilitate a stronger evolution in the inner region of galaxies as they are expected to undergo an increased amount of accretion and interaction with satellites \citep[][and references therein]{Conselice2014}. In our study also, in absolute terms, the increase in the inner density is observed to be more prominent in the case of the normal disc sample. Thus, environment appears to have strongly affected bulge growth over the past $\sim$8 Gyrs.

Overall, we have found that both internal and external mechanisms are involved in disc and bulge evolution. External processes, in the form of minor mergers and accretion, appear to be playing a more effective role in growing classical bulges in relatively denser environments.

Examining the inner region properties through both parametric and non-parametric measures provides considerable insight into the relative role of the processes involved in disc evolution. Further understanding can perhaps be obtained from studies in other wavelengths, especially the infrared region, in which dust effects are minimised. 

\section*{Acknowledgements}
We are thankful to Christopher J. Conselice, Michael Blanton and Knud Jahnke for providing useful inputs. SS is thankful to receive Senior research fellowship (09/045(0972)/2010-EMR-I) from Council of Scientific and Industrial Research (CSIR), India. SS also acknowledges the grant received as a DGDF-2014 project student at ESO, Chile. SS is thankful to Ranjeev Misra, Ajit Kembhavi and Yogesh Wadadekar for useful discussions. Lastly, SS would like to thank IUCAA, India and ESO, Chile for the local hospitality provided during the course of this work. We are highly grateful to the referee for providing detailed and insightful comments, which has improved the presentation of the work. 


\bibliographystyle{mn2e}
\bibliography{paper_final}
\label{lastpage}
\end{document}